\documentclass[aps,prl,lengthcheck,superscriptaddress]{revtex4-1}
\usepackage{amsmath,amssymb}
\usepackage{amsfonts}
\usepackage{dcolumn}
\usepackage{braket, bm}
\usepackage{graphicx,color}
\usepackage[capitalize]{cleveref}
\begin{document}
\title{Competition and interplay between topology and quasi-periodic disorder in Thouless pumping of ultracold atoms}

\author{Shuta Nakajima}
\altaffiliation{Electronic address: shuta@scphys.kyoto-u.ac.jp}
\affiliation{Department of Physics, Graduate School of Science, Kyoto University, Kyoto 606-8502, Japan}
\affiliation{The Hakubi Center for Advanced Research, Kyoto University, Kyoto 606-8502, Japan}
\author{Nobuyuki Takei}
\affiliation{Department of Physics, Graduate School of Science, Kyoto University, Kyoto 606-8502, Japan}
\date{\today}
\author{Keita Sakuma}
\affiliation{Department of Physics, Graduate School of Science, Kyoto University, Kyoto 606-8502, Japan}
\author{Yoshihito Kuno}
\affiliation{Department of Physics, Graduate School of Science, Kyoto University, Kyoto 606-8502, Japan}
\affiliation{Department of Physics, University of Tsukuba, Tsukuba, Ibaraki 305-8571, Japan}
\author{Pasquale Marra}
\affiliation{Graduate School of Mathematical Sciences, The University of Tokyo, Komaba, Tokyo, 153-8914, Japan}
\affiliation{Department of Physics, and Research and Education Center for Natural Sciences, Keio University, Hiyoshi, Kanagawa, 223-8521, Japan}
\author{Yoshiro Takahashi}
\affiliation{Department of Physics, Graduate School of Science, Kyoto University, Kyoto 606-8502, Japan}

\begin{abstract}
Robustness against perturbations lies at the heart of topological phenomena. 
If, however, a perturbation such as disorder becomes dominant, 
it may cause a topological phase transition between topologically non-trivial and trivial phases. 
Here we experimentally reveal the competition and interplay between topology and quasi-periodic disorder in a Thouless pump realized with ultracold atoms in an optical lattice, 
by creating a quasi-periodic potential from weak to strong regimes in a controllable manner. 
We demonstrate a disorder-induced pumping in which the presence of quasi-periodic disorder can induce a non-trivial pump for a specific pumping sequence, 
while no pump is observed in the clean limit. 
Our highly controllable system,
which can also straightforwardly incorporate interatomic interaction, 
could be a unique platform for studying various disorder-related novel effects in a wide range of topological quantum phenomena.

\end{abstract}
\maketitle


While arbitrarily weak random disorder drives non-interacting particles in one or two dimensions into a localized state,
known as the Anderson localization~\cite{50AL}, 
topological quantum phenomena are robust against weak disorder~\cite{Niu84,Niu85}, i.e., topological phenomena avoid to be Anderson-localized.
Robustness to disorder is an important property defining the topological states,
the examples of which are the classification of three-dimensional ${\mathbb{Z}}_{2}$ topological insulators into strong and weak~\cite{Kobayashi,Bryan,Yamakage}.
More recent investigations focus not only on the robustness but also on the novel interplay between disorder and topology, exhibiting complex but interesting phenomena~\cite{Li2009,McGinley}.
A striking example is the topological Anderson insulator~(TAI)~\cite{Li2009,Shen}.
Here, the introduction of a disorder potential into a two-dimensional system is essential to create a topologically non-trivial phase:
Counter-intuitively, while the system is topologically trivial in the clean limit, 
disorder acts as a protection, instead of suppression, of the topological invariant, such as the Chern number.
Two recent experiments~\cite{Meier,Stutzer2018} studied a non-interacting bosonic or photonic waveguide system of TAI 
using a Bragg-coupled one-dimensional array of Bose-Einstein condensates in a momentum-space lattice~\cite{Meier} 
or a two-dimensional array of evanescently coupled helical waveguides described by the paraxial wave equation~\cite{Stutzer2018}, 
revealing the disorder-induced chiral displacement or edge states, respectively.
Moreover, several topological properties of edge-state propagation in a photonic TAI have been recently
demonstrated in a two-dimensional disordered gyromagnetic photonic crystal in the microwave regime~\cite{Liu2020}.
Further theoretical works on TAI envision rich varieties of disorder-induced topological phenomena like an anomalous Floquet Anderson insulator~\cite{Titum},
a Majorana topological phase~\cite{McGinley},
a Dirac-semi-metal~\cite{Sriluckshmy},
and ${\mathbb{Z}}_{2}$ topological insulators~\cite{Kobayashi, Mondragon-Shem, Yamakage}.

Historically, the robustness of the topological invariant of Chern number against perturbations was first discussed by Niu and Thouless~\cite{Niu84,Niu85} who investigated the effects of spatial potential disorder and inter-particle interaction on a topological Thouless pump~\cite{Thouless}. 
This Thouless pump is quantum transport of a fermionic gas in a one-dimensional periodic potential driven in an adiabatic cycle.
The charge pumped per cycle is equal to the Chern number defined over a two-dimensional Brillouin zone with one spatial and one temporal dimension, and thus it can be regarded as a (1+1)-dimensional counterpart of the quantum Hall effect. 
They derived an effective Chern number even under perturbations by introducing the twisted boundary conditions. 
As a result, they revealed that the amount of pumping corresponding to the Chern number does not change unless the gap between the associated energy bands is closed.
The experimental realization of Thouless pump in a clean system using ultracold atoms in optical lattices~\cite{Nakajima, Lohse, Schweizer2016, Lohse2018} and photonic waveguides~\cite{Kraus, Zilberberg2018} triggers various interesting theoretical investigations on the effect of the perturbation
like on-site static~\cite{Qin2016,Wauters,Imura_2018,Kuno} 
and dynamic~\cite{Titum} potentials as well as
interatomic interaction~\cite{Nakagawa,Hayward,Stenzel,Mei_2019}.

\begin{figure}[tbp]
 \includegraphics[width=8.5cm,clip]{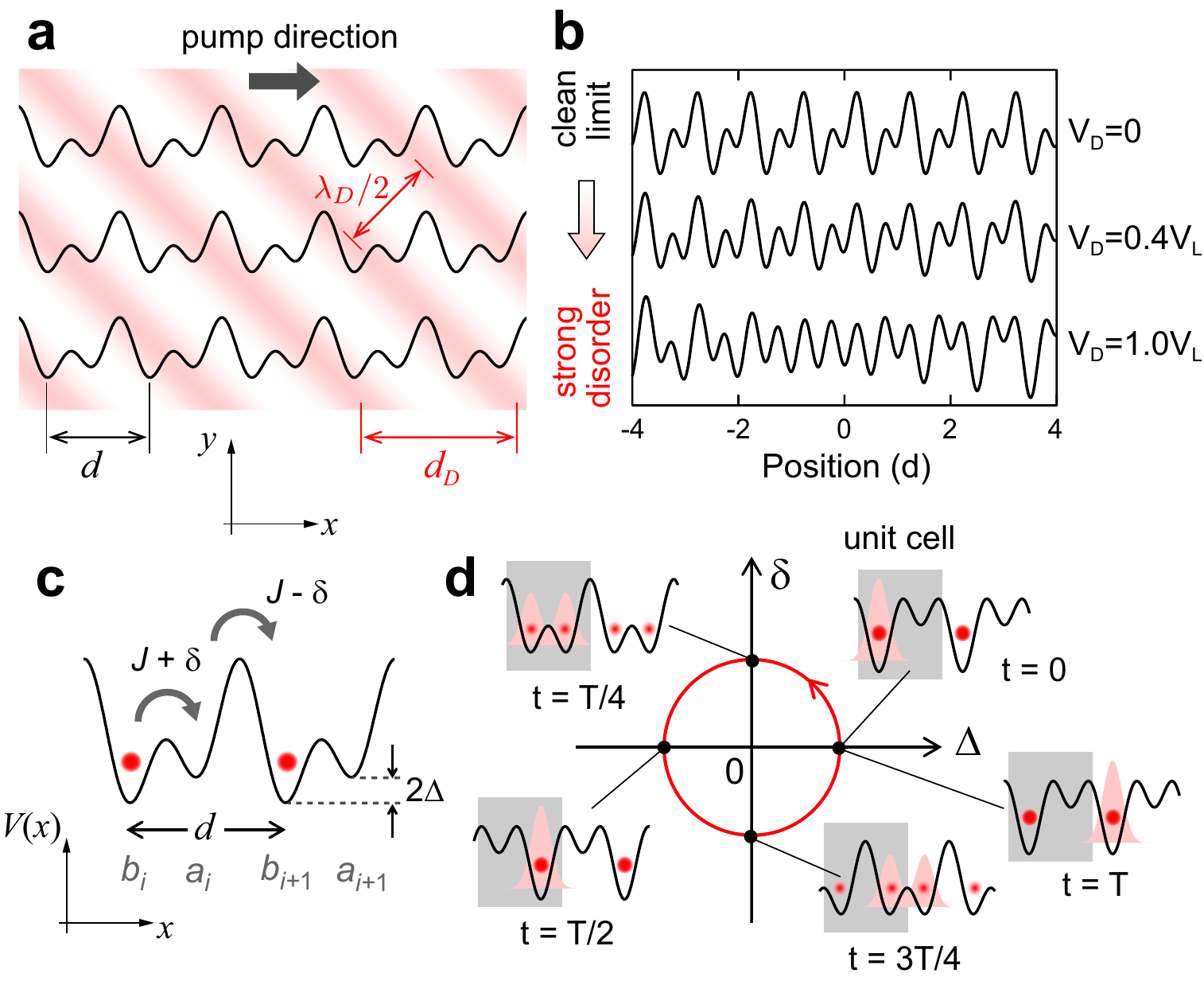}
\caption{
{\bf Rice-Mele model.}
\textbf{a},~Lattice configuration of a 2D array of continuous Rice-Mele (cRM) lattices (black) and disorder lattice (red) tilted by $45^\circ$ with respect to the pump direction $x$.
\textbf{b},~cRM lattice with $V_S=V_L$ combined with a disorder lattice at $\lambda_D=798\,\mathrm{nm}$. From top to bottom, the quasi-periodic disorder strengths $V_D$ are $0$ (the clean limit), $0.4V_L$, and $1.0V_L$, respectively.
\textbf{c},~Schematic of the cRM model in the clean limit.
\textbf{d},~A pumping cycle of the cRM sequence with a period of $T$, pictorially sketched in $\delta$-$\Delta$ space.
The gray shaded box represents a unit cell, and the pink shaded packet indicates the wavefunction of a particular atom.
}
\label{fig:RM}
\end{figure}

Here, employing a Thouless pump realized with ultracold fermions in a dynamical optical lattice, we experimentally reveal the competition and interplay 
between topology and quasi-periodic disorder by generating a controllable quasi-periodic potential from weak to strong regimes. 
As the highlight, by choosing pump parameters appropriately, we successfully demonstrate a disorder-induced pumping, 
i.e., a trivial phase with no pump in the clean limit is driven to a nontrivial phase with finite pump,
due to the presence of quasi-periodic disorder, and is driven back to a trivial phase at larger disorder strengths.
This phenomenon is the (1+1) dimensional analogue of TAI originally discussed 
in a two-dimensional system~\cite{Li2009}. 
Moreover, our experimental observations not only demonstrate the realization of disorder-induced pump, but also
quantitatively reveals the degree of robustness and breakdown of the Thouless pump against quasi-periodic disorder.
On the one hand, 
the pumped charge does not change even at disorder strengths comparable to the Anderson-localization transition point 
and to the minimum band gap in the clean limit, while on the other hand,
the pumped charge drastically decreases when the disorder strength exceeds the threshold value determined by the pump parameters.
In addition, our further measurement suggests that the gap closes at the threshold of the disorder strength, 
indicating that the quasi-periodic disorder induces a topological phase transition from topologically non-trivial to trivial phases.


\subsection*{Experimental setup of Thouless pumping with quasi-periodic disorder}
We experimentally implement Thouless pumping in the presence 
of quasi-periodic disorder with precisely controllable strength,
using an ultracold non-interacting Fermi gas of ytterbium ($^{171}$Yb) atoms in a two-dimensional (2D) array of continuous Rice-Mele (cRM) lattices (see Methods)~\cite{Nakajima}.
Each lattice is a dynamically controlled optical superlattice which consists of a dynamical interferometric lattice with the lattice constant $d=532$~nm (``long lattice'') and a stationary lattice with the constant $d/2$ (``short lattice'').
We superimpose another optical lattice at wavelength $\lambda_D$ on it, which is tilted by $45^\circ$ with respect to the pumping direction $x$ as depicted in Fig. 1a.
Hereafter we refer to this lattice as the ``disorder lattice''.
Due to strong confinement along the $y$ and $z$ axes provided by other optical lattices, the following time-dependent one-dimensional (1D) superlattice is created:
\begin{eqnarray}
V(x,t) &=& -V_S\cos^2\left(\frac{2\pi x}{d}\right)-V_L\cos^2\left(\frac{\pi x}{d}-\phi(t)\right)\nonumber \\ 
&&-V_D\cos^2\left(\frac{\pi x}{d_D}+\frac{\alpha}{2} \right),
\label{eq:lattice}
\end{eqnarray}
where $V_S$, $V_L$, and $V_D$ correspond to the depth of the short, long, and disorder lattices, respectively.
$d_D=\lambda_D/\sqrt{2}$ is the lattice constant of the disorder lattice along the pumping direction. When $d_D \neq d$, i.e. $\lambda_D \neq 752$~nm, the quasi-periodic disorder is produced, and its strength can be controlled by adjusting the depth $V_D$ as shown in Fig. 1b. 
$\phi$ is the phase difference between the short and long lattices and $\alpha$ is the phase difference between the short and disorder lattices.
Note that this phase $\alpha$ takes a different value for different 1D superlattice due to our configuration in Fig. 1a (see Methods). 
Hereafter, we use the lattice constant $d$ as the unit of length and the recoil energy $E_R=h^2/(8md^2)$ as the unit of energy, 
where $h$ denotes Planck's constant and $m$ is the atomic mass of $^{174}$Yb.
We load the ytterbium atoms into the 2D array of 1D cRM lattices
in the staggered phase ($\phi=0$),
first by ramping-up the long lattice followed by the simultaneous ramping-up of short and disorder lattices adiabatically, ensuring that they occupy the lowest energy band, and slowly sweep $\phi$ over time.
The lattice potential returns to its initial configuration whenever $\phi$ changes by $\pi$, thus completing a pumping cycle.
The typical length of a 1D superlattice is about 12 unit cells around the center of the atomic cloud.

Our Thouless pump under quasi-periodic disorder can be approximately described by the tight-binding Rice-Mele (tRM) model~\cite{RM,Atala} with on-site quasi-periodic disorder: 
\begin{eqnarray}\label{eq:RM}
\mathcal{\hat{H}}&=& \sum_i\left(-(J+\delta(t)) \hat{a}_i^\dagger \hat{b}_i -(J-\delta(t))\hat{a}_i^\dagger\hat{b}_{i+1}+{\rm h.c.}\right. \nonumber \\
&&\left.+\Delta(t) (\hat{a}_i^\dagger \hat{a}_i-\hat{b}_i^\dagger \hat{b}_i)
+\Delta_D^{a,i}\hat{a}_i^\dagger \hat{a}_i + \Delta_D^{b,i}\hat{b}_i^\dagger \hat{b}_i\right),
\end{eqnarray}
where $\hat{a}_i^\dagger$ ($\hat{a}_i$) and $\hat{b}_i^\dagger$ ($\hat{b}_i$) 
are fermionic creation (annihilation) operators in the two sublattices of the $i$-th unit cell,
$J\pm \delta(t)$ is the tunneling amplitude within and between unit cells, $\Delta(t)$ denotes a staggered on-site energy offset ($\max |\Delta(t)| \equiv \Delta_0$),
and $\Delta_D^{a(b),i}$ is the on-site quasi-periodic disorder of sublattices $a$ ($b$) of the $i$-th unit cell
($\displaystyle\max_{\it i} |\Delta_D^{a(b),i}| \equiv \Delta_D/2$, see Supplementary Information~S1 for details).
Figure~\ref{fig:RM}c represents this system in the corresponding cRM model in the clean limit $\Delta_D^{a(b),i}=0$ (or $V_D=0$).
By sweeping the long lattice phase $\phi(t)$, dynamical parameters $\delta(t)$ and $\Delta(t)$ change adiabatically and draw a closed trajectory in a $\delta-\Delta$ space (Fig.~\ref{fig:RM}d). 
The parameters $(J,\delta(t),\Delta(t))$ are determined by matching 
the band structures of the tRM and cRM models corresponding to the points ($\phi=0,\pi/2,\pi,3\pi/2$) of the actual pumping sequences. 
For example, when $(V_{L},V_{S})=(20,14)E_R$ and for $\phi=0$, 
we get $(J,\delta,\Delta)=(0.861,0,6.45)E_R$, whereas for $\phi=\pi/2$, 
we get $(J,\delta,\Delta)=(0.861,0.852,0)E_R$.
We then swipe $(\delta(t),\Delta(t))$ along the circular path in Fig.~\ref{fig:RM}d.
Such mapping does preserve the topological properties of the system.
Note that the actual experiment is only approximately described by the tRM model,
so that we use the circular trajectory only as a pictorial description of the pumping sequences~\cite{Nakajima}.


\begin{figure}[tbp]
	\begin{center}
 \includegraphics[width=8.0cm,clip]{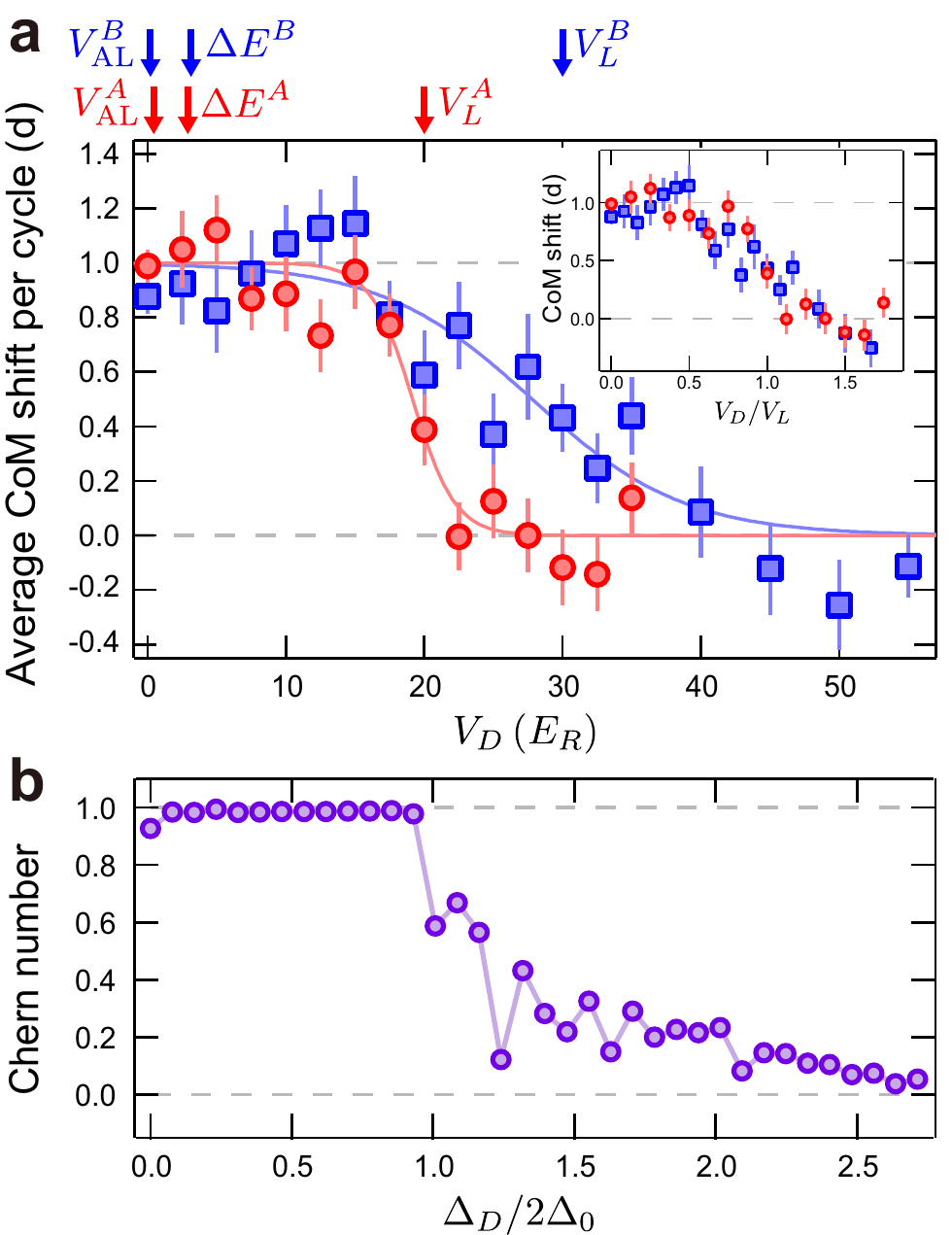}
\caption{
\textbf{Breakdown of the Thouless pump under quasi-periodic disorder.}
\textbf{a},~The CoM shift per cycle averaged after three cycles, plotted as a function of the disorder lattice depth $V_D$ for  cRM lattice with $(V_L, V_S)=(20, 14)E_R$ (set A, red circles) and $(V_L, V_S)=(30, 20)E_R$ (set B, blue squares). The vertical red (blue) arrows indicate three energies in the pump; from left to right, 
$V_{\rm AL}^A=0.3E_R$ ($V_{\rm AL}^B=0.07E_R$) AL transition point $V_{\rm AL}$, 
$\Delta E^A=3.25E_R$ ($\Delta E^B=3.38E_R$) minimum band gap of the bare cRM model $\Delta E$, and 
$V_L^A = 20E_R$ ($V_L^B = 30E_R$) the depth of long lattice corresponding to 
the maximum onsite offset $2\Delta_0$.
The inset shows the plots of the same data
as a function of a normalized disorder strength $V_D/V_L$ for each parameter set.
The error bars denote the $1\sigma$ confidence bound derived from more than forty CoM measurements.
\textbf{b},~Numerical calculation of the Chern number in the presence of quasi-periodic disorder based on
our tight-binding model.
The result was obtained by averaging over the quasi-periodic disorder phase $\alpha$.
}
\label{fig:dis_dep} 
\end{center}
\end{figure}

\subsection*{Effect of quasi-periodic disorder on Thouless pump}
We begin by considering the effects of quasi-periodic disorder added to our cRM version of the Thouless pump. 
While Niu and Thouless showed that the pump is robust against disorder as long as the band gap remains open,
it is yet unclear how much quasi-periodic disorder is needed in order to close the gap.
While infinitely small random disorder induces localization in 1D systems,
a finite quasi-periodic disorder is needed in order to induce localization.
In our system, there are three energy scales which might be relevant for gap-closing:
1) the Anderson-localization (AL) transition point $V_{\rm AL}$, 
2) the minimum band gap $\Delta E$ ($\sim$ maximum $J$) in the pumping cycle of the bare cRM lattice in the clean limit, and 
3) the maximum on-site offset $\Delta_0$ of the staggered potential. 
By performing the experiments we can answer which energy scale is truly relevant.
Figure~\ref{fig:dis_dep}a shows the results of cRM pump under the quasi-periodic disorder potential with $\lambda_D=798$~nm for two sets of superlattice parameters, 
$(V_L,V_S)=(20, 14)E_{\rm R}$ (set A, red circles)
and $(V_L,V_S)=(30, 20)E_{\rm R}$ (set B, blue squares) as a function of the disorder strength $V_D$.
The pumped charge per cycle was obtained by evaluating the shift of the center of mass (CoM) of the atomic cloud, which corresponds to the Chern number~\cite{Nakajima}. 
The vertical red (blue) arrows in Fig.~\ref{fig:dis_dep}a 
indicate the above-mentioned three energy scales for the superlattice parameters of the set A (B).
Here, the arrows at $\sim 0.1~E_R$ with $V_{\rm AL}^{A(B)}$
and those at $\sim 3~E_R$ with $\Delta E^{A(B)}$
correspond to the AL transition points $V_{\rm AL}$ (see Supplementary Information~S2) and the minimum gap $\Delta E$ for the superlattice parameter A (B), respectively. 
The arrows with $V_L^A$ and $V_L^B$ at $20~E_R$ and $30~E_R$ represent the lattice-depth of $V_L$ for the parameter set A and B, respectively. 
The measured results demonstrate that the Thouless pump is robust against disorder strengths
around the AL transition point $V_{\rm AL}$ and minimum gap $\Delta E$, 
and it breaks only when the disorder strength becomes comparable with $V_L$.
In our cRM lattice, the value of $V_L$ is comparable to the maximum on-site offset $2\Delta_0$ in the tRM model. By normalizing the horizontal axes of Fig.~\ref{fig:dis_dep}a with the values of $V_L$ for each lattice setting, 
we can evaluate the pumping behavior in terms of the adimensional disorder strength $V_D /V_L$, as shown in inset of Fig.~\ref{fig:dis_dep}a.
Indeed, the pumping is suppressed at the disorder strength of $V_D/V_L \sim 1$.
Although non-adiabatic processes shift the suppression point toward weaker disorder,
they result in only a small underestimation of the suppression point and do not play the dominant role~\cite{Hayward2020}. Our experiment shows that the topological transition occurs well in the regime where all states are Anderson-localized and is not regarded as a delocalization-localization transition of instantaneous Hamiltonian eigenstates~\cite{Hayward2020}. Instead, the observed breakdown of quantized pumping is regarded as manifestation of a delocalization-localization transition of Floquet eigenstates~ \cite{Wauters}.

To explain the experimental results in Fig.~\ref{fig:dis_dep}a, 
we numerically calculate the Chern number of the tRM model with on-site quasi-periodic disorder. 
In general, the Chern number corresponds to the total pumped charge per pumping cycle 
if the band gap is kept open during the pumping cycle \cite{Asboth}.
Here, due to the presence of quasi-periodic disorder, the tRM model does not have translational invariance: 
Hence,
the conventional definition of Chern number in momentum space cannot be applied. 
Instead, we applied the dimensional extension for the tRM model and introduced twisted phase boundary condition, and 
by employing the coupling matrix method \cite{YFZhang,Sriluckshmy,Kuno} we calculate a modified Chern number, 
i.e., extracted a quantity related to the total pumped charge from the real space disorded Hamiltonian (see Supplementary Information~S3).

Figure~\ref{fig:dis_dep}b shows the results of numerical calculations of the Chern number for a superlattice with 20 unit cells. 
In our numerical result, the Chern number takes non-integer values around the transition point (moderate disorder regime).
This has two reasons.
The first reason is that the evaluated Chern number corresponds to the average over the quasi-periodic disorder phase $\alpha$.
Because the phase $\alpha$ varies among 1D superlattices, 
the transition strength $V_D$ at which the Chern number changes from one to zero is different for a different lattice.
In the actual experiment, all the 1D lattices are measured simultaneously, 
so that the phase dependency is averaged out.
The second reason is that the gap fluctuates by varying the twisted phases at the boundary in the strong disorder regime \cite{YFZhang,Kuno}.
The numerical calculations and measured results are in reasonable agreement, and also consistent with the numerical results in Refs.~\cite{Wauters} and \cite{Hayward2020}.
In general, we expect deviations between numerical calculations and experiments due to various factors: 
(I) effects of residual non-adiabaticity in the experiment, 
(II) absence of disorder in the hopping terms of the tRM model,
(III) effects of the trapping potential not considered in tRM model. 
These factors are also relevant for the experimental results and numerical calculations presented in Fig.~\ref{fig:DITCP}.

\begin{figure}[tbp]
	\begin{center}
 \includegraphics[width=7.5cm,clip]{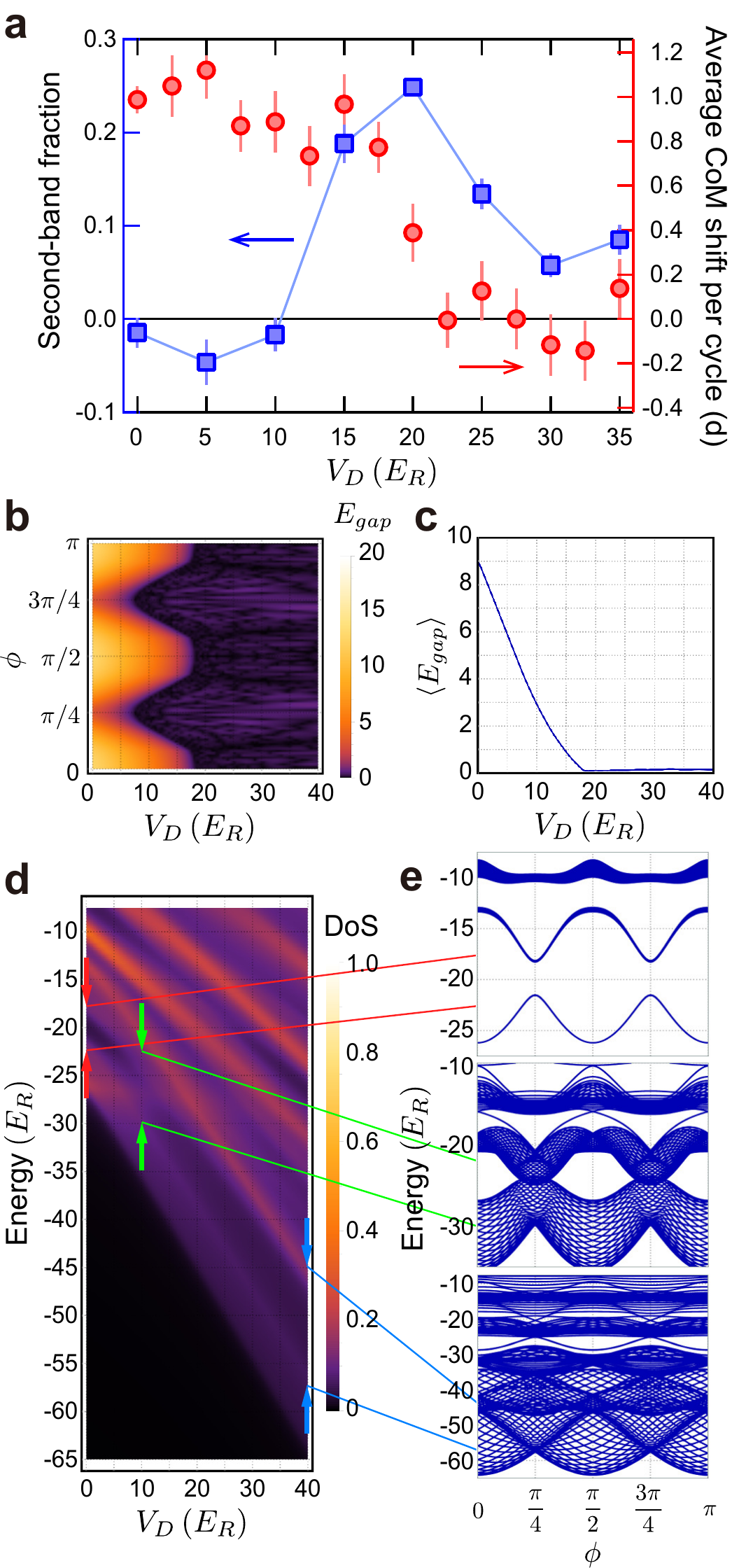}
\caption{
\textbf{Gap closing and opening induced by the quasi-periodic disorder.} 
\textbf{a},~The second-band fraction measured after three pumping cycles for cRM lattice with the parameter set A $(V_L, V_S)=(20, 14)E_R$ plotted as a function of disorder-lattice depth, represented by blue squares.
The error bars denote the $1\sigma$ confidence bound derived from more than ten band-mapping measurements.
As a reference, we show the corresponding data of charge pumping by red circles, which are the same as the red circles in Fig.~\ref{fig:dis_dep}a.
The population takes a maximum around the region of the breakdown of the pumping.
The negative values at weak disorder are due to the fluctuation of the number of atoms.
\textbf{b},~2D plot of energy gap as functions of superlattice phase and the disorder strength $V_D$ 
calculated based on the continuous model under half filling (one atom per unit cell) with the parameter set A.
\textbf{c},~Energy gap averaged over $\phi$ as a function of the disorder strength $V_D$. 
\textbf{d},~DoS as a function of the disorder lattice strength
calculated using a continuous model (with arbitrary units) averaged over the phase $\phi$.
\textbf{e},~Band structures with different disorder lattice strengths (from top to bottom for $V_D=0, 10$, and $40~E_R$)
calculated using a continuous model.
}
\label{fig:dis_BM} 
\end{center}
\end{figure}

It is important to clarify whether the observed suppression of the charge pumping 
is the manifestation of a topological phase transition or not.
A fundamental property which is useful to identify a topological phase transition is the behavior of the excited band population.
This has been the key to reveal the topological phase transition of the Haldane model~\cite{Jotzu14}, 
in which the population of the second band increases at the topological transition point, where the gap closes, 
due to the Landau-Zener transition during the Bloch oscillation.
Here, we investigate the gap closing between the first and the second bands by the presence of the quasi-periodic disorder with a band mapping technique (see Supplementary Information~S5).
Figure~\ref{fig:dis_BM}a shows the fraction of atoms excited to the second band, with respect to the total number of atoms, after three cycles of cRM pumping with the parameter set A
$(V_L, V_S) = (20, 14)E_R$.
We evaluate the population of the second Brillouin zone of the basic cRM lattice.
To this end, we remove the disorder lattice by adiabatically ramping down its strength $V_D$ 
before the band-mapping measurements, so that the fraction of atoms in the second-band under quasi-periodic disorder
 is adiabatically transferred to the corresponding one in the clean limit.
The experimental result shows that the excitation to the second band initially increases
as disorder increases and reaches the maximum at $V_D \sim V_L$ and then decreases as the disorder further increases. 
Such increase of the excitation to the second band around $V_D \sim V_L$ suggests that the gap initially opened in a clean limit is closed around $V_D \sim V_L$ and becomes open again.
The band gap changes dynamically during the pump cycle, and a Landau-Zener-like non-adiabatic transition could occur at the gap closing point. 
Note that, for strong disorder, the gap may close during the removal of the disorder lattice, which is performed at $\phi=0$ (see Fig. 3b).
However, if the gap remains closed even in the region of $ V_D/V_L \gg 1 $, 
it is not expected to observe a decrease 
in the second-band fraction in the strong disorder region, as it is instead observed in the experiment.

We check this gap closing and opening around $ V_D \sim V_L $ numerically.
Figure~\ref{fig:dis_BM}b shows a numerical calculation of the energy gap $E_{\mathrm{gap}}$ for the set A as functions of $V_D$ and $\phi$ 
based on a continuous model\cite{Das2019,Marra2020} in the presence of a quasi-periodic superlattice
with half filling and a total length of 198~$d$ (see Supplementary Information~S4 for details).
The energy gap averaged over $\phi$ becomes zero around $V_D \sim 20E_R$ (see Fig.~\ref{fig:dis_BM}c).
It qualitatively supports our experimental observations that the pumping is suppressed at the disorder strength of $V_D/V_L \sim 1$.
The disappearance of the bulk energy gap also can be seen in the density of states (DoS) calculation and the band calculation (Fig.~\ref{fig:dis_BM}d and e).
Here, as increasing $V_D$, some in-gap states are formed and the bulk band gap observed in the clean limit gradually disappears.
On the other hand, the numerical calculations do not show such re-opening of the gap.
However, the energy difference between the ground vibrational level (the local first band) and the first excited vibrational level (the local second band) within one lattice site would become large in the strong (deep) disorder lattice regime. This large local (on-site) energy gap in large $V_D$ suppresses again the excitation to the second band during pumping sequence as is observed in our experiment. Our consideration of the sliding and disorder lattices (see Supplementary Information S6 and later in the main text)
indicates that the gap should be open locally on each site for $V_D \gg V_L$.
We also numerically confirmed that the gap closing and reopening can be seen more clearly in the case of shallower lattice parameter sets
(see Supplementary Information~S4 for details).
Consequently, the observed pumping suppression due to the quasi-periodic disorder
should indicate a topological phase transition in which the non-trivial Chern number changes into a trivial one via gap closing as the disorder strength increases.


\begin{figure}[tbp]
 \includegraphics[width=8.3cm,clip]{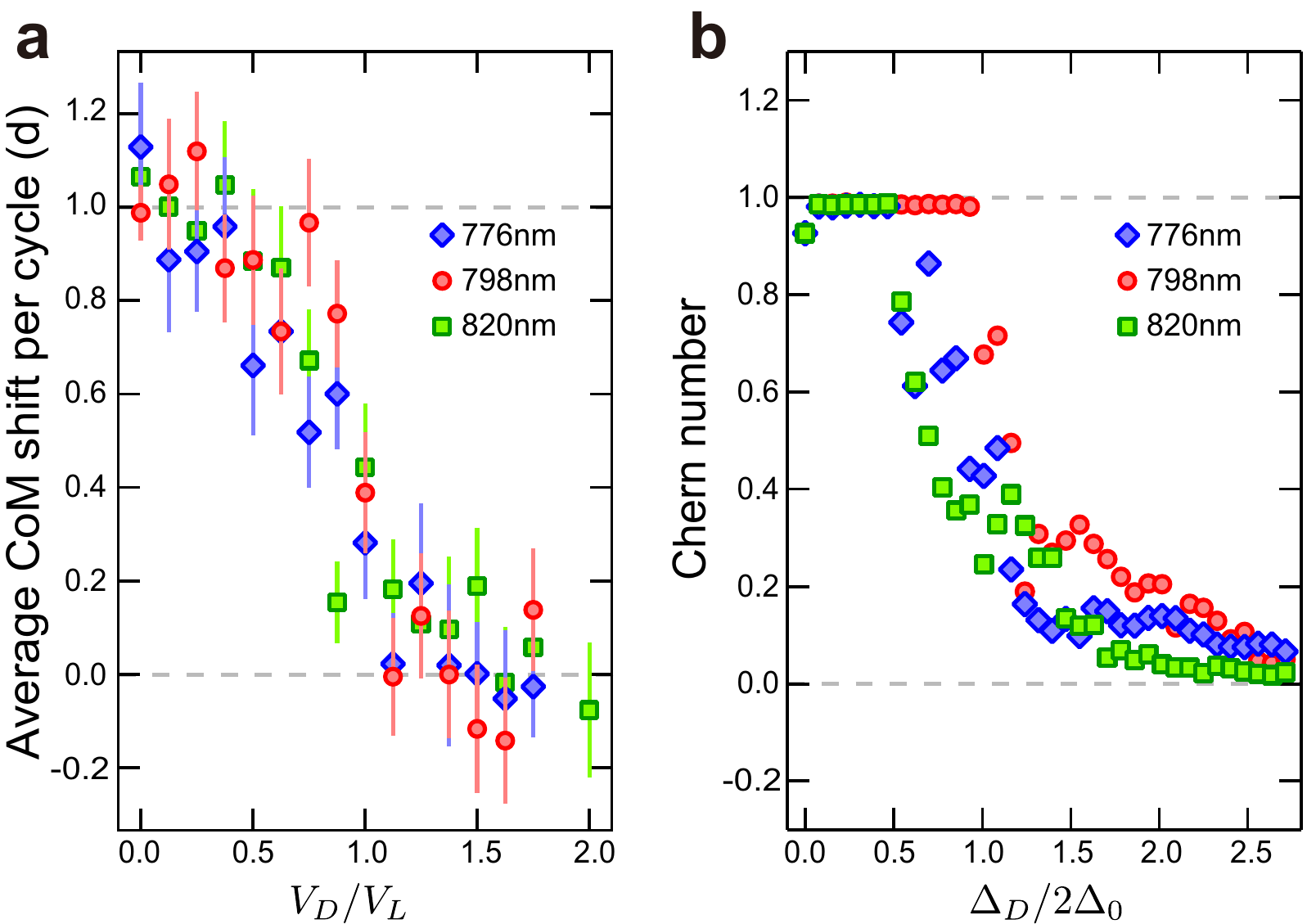}
\caption{
\textbf{Dependence of pumping suppression on the wavelength of the quasi-periodic disorder lattice.}
\textbf{a},~The CoM shift per cycle averaged after three cycles
plotted as a function of the normalized depth of the disorder lattice with the wavelength of $\lambda_D=776$~nm (blue diamond), 798~nm (red circle), and 820~nm (green square)
for cRM lattice with $(V_L, V_S)=(20, 14)E_R$.
The error bars denote the $1\sigma$ confidence bound derived from more than forty CoM measurements.
\textbf{b},~Numerical calculation of the wavelength dependence of the Chern number, plotted as a function of the normalized disorder strength in the tight-binding model. 
The symbols are the same as those in \textbf{a}. 
}
\label{fig:WL_dep}
\end{figure}

\begin{figure*}[tbp]
	\begin{center}
 \includegraphics[width=16cm,clip]{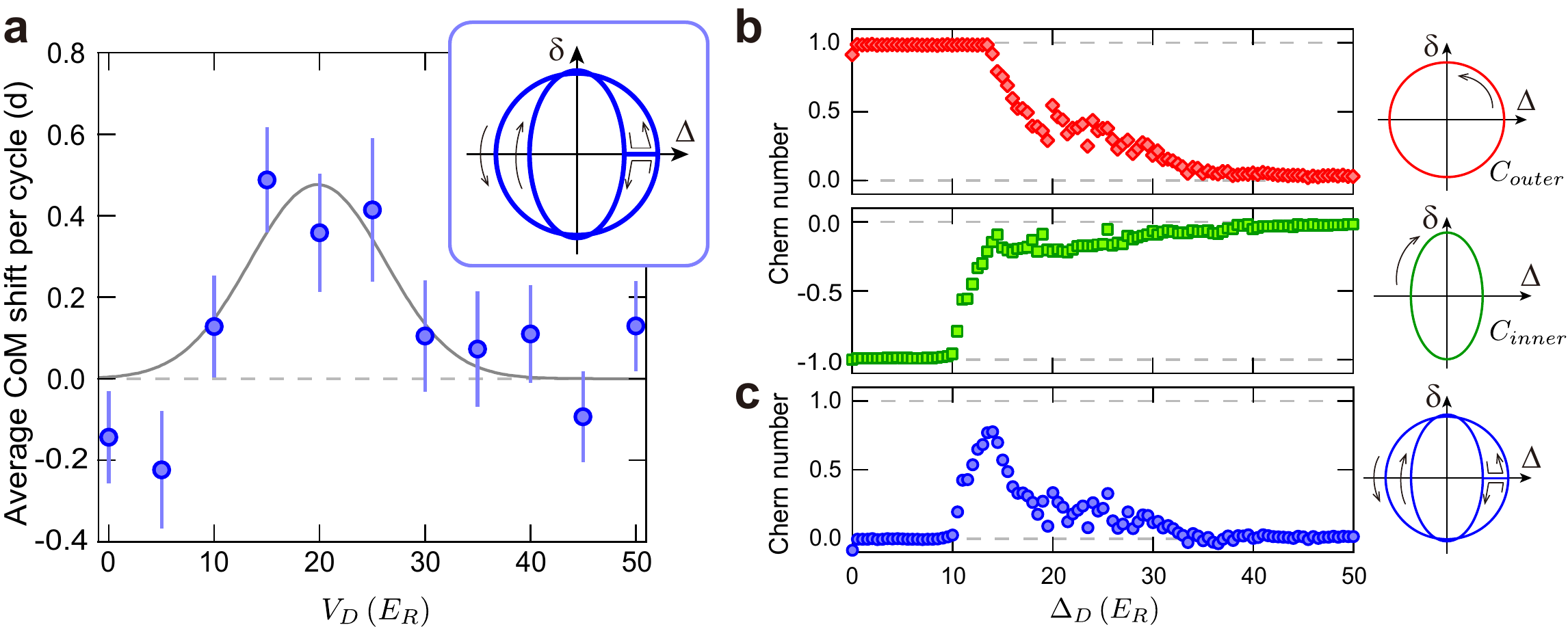}
\caption{
{\bf Disorder-induced pumping.}
\textbf{a},~The CoM shift per cycle averaged after three cycles plotted as a function of the quasi-periodic disorder lattice depth $V_D$.
The gray line is a Gaussian fit and used as a guide to eye.
The inset shows pictorially the pumping sequence in which the outer and inner loops are expressed by the cRM lattice parameters $(V_L,V_S)=(36,24)E_R$ and $(15,10)E_R$, respectively, and connected by changing the lattice parameters at the staggered phase $(\phi=0)$.
The pumping direction of the inner loop is opposite to that of the outer one.
The error bars denote the $1\sigma$ confidence bound derived from more than forty CoM measurements.
\textbf{b},~Numerical calculations of Chern number for the outer (top) and inner (bottom) trajectories, shown on the right side, as a function of $\Delta_D$.
\textbf{c},~The sum of the Chern number for two trajectories in \textbf{b}.
}
\label{fig:DITCP} 
\end{center}
\end{figure*}

We also examined the dependence of the pumping suppression on the wavelength of the disorder lattice $\lambda_D$.
Figure~\ref{fig:WL_dep}a shows the pumping amounts measured at the quasi-periodic disorder wavelengths $\lambda_D=776$~nm (red diamond), 798~nm (blue circle), and 820~nm (green square).
The data at 798~nm are the same as those with the set A in Fig.~\ref{fig:dis_dep}.
There is no clear difference among the results for those different wavelengths.
Our numerical calculation with a tRM model supports the tendency of the measured results (see Fig.~\ref{fig:WL_dep}b).
As one can see, the charge pumping starts to decrease from one at a certain critical disorder strength, which depends on the wavelength. However, the pumping is largely suppressed at almost the same value $V_{D}\gtrsim 20 E_R$ for all three wavelengths.

Although it is difficult to fully describe the pump in the strong disorder regime with such a tRM model, 
intuitively we can understand why the pump is suppressed at $ V_D \sim V_L $ regardless of the 
wavelength of the disorder lattice.
Our cRM pump is topologically equivalent to a simple sliding lattice, namely, the second term of Eq.(\ref{eq:lattice}) describing the long lattice only~\cite{Nakajima}. 
Thus we can capture the essence of the results by considering a sliding lattice superimposed with a disorder lattice.
A simple calculation shows that, in the case of $V_D < V_L$, the sliding long lattice is dominant, so that the total combined lattice slides similarly.
On the other hand, when $ V_D > V_L $, the minima of the total lattice do not slide
(see Supplementary Information~S6).
From this point of view, the former corresponds to a quantized pumping, whereas the latter gives no pump.


\subsection*{Disorder-induced pumping}
The introduction of strong quasi-periodic disorder suppresses the topological charge pumping as described above.
However, the interplay between topology and quasi-periodic disorder could lead to more counter-intuitive phenomena like TAI~\cite{Li2009,Shen}.
Here, we demonstrate the novel phenomenon of disorder-induced pumping, realized with a newly configured pumping sequence. 
We designed a pumping sequence in which the charge pumping is implemented in three stages:
using the lattice parameters $(V_L,V_S)=(36,24)E_R$ in the first stage, 
then using different parameters $(V_L,V_S)=(15,10)E_R$ with pumping in the reverse direction in the second stage, 
and finally using the lattice parameters of the first stage to obtain a closed pumping cycle.
This sequence is depicted in a pictorial way in Fig.~\ref{fig:DITCP}a, 
in the tRM model.
In a clean system, the gap closes only at $(\delta,\Delta)=(0,0)$ and is open in the other region, so that the outer loop yields the Chern number of $\mathcal{C}_{\rm outer}=+1$ and the inner loop $\mathcal{C}_{\rm inner}=-1$.
Thus the overall pumping amount in this sequence can be given by $\mathcal{C}_{\rm outer}+\mathcal{C}_{\rm inner}=0$. 
Theoretically, if the two loops are far apart, a quantized pump is expected under moderate disorder (See Supplementary Information S3).
We added the disorder lattice to this basic sequence and measured the pumping amount per one cycle shown in Fig.~\ref{fig:DITCP}b.
The pumping amount was obtained by evaluating the CoM shift after three cycles for each disorder strength.
At zero disorder strength, the pumping amount is zero as discussed above.
However, as the disorder strength increases, the pumping amount takes a finite value.
Subsequently, the pumping amount reaches its maximum value and then reduces to zero.
This clearly demonstrates the observation of disorder-induced pumping in the Thouless pump system.

Our finding can be explained intuitively in the following manner.
As discussed in Figs.~\ref{fig:dis_dep}a and b, the pumping is suppressed 
at disorder strengths of the order of $V_L$.
Namely, in Fig.~\ref{fig:DITCP}a, the pumping is expected to be suppressed around $V_D = 36~E_R$ and $15~E_R$ in the outer and the inner trajectories, respectively.
This indicates that in the intermediate region  $15E_R\lesssim V_D \lesssim 36E_R$
the pumping during the inner and outer trajectories do not cancel each other, resulting in a non-zero total pumping.

We qualitatively reproduced this behaviour by numerical calculations as shown in Fig.~\ref{fig:DITCP}b.
For each circular trajectory, we calculated the Chern number by using the quasi-periodic disordered Harper-Hofstadter-Hatsugai model (see Supplementary Information~S3 for details). 
For all data sets, we averaged over 60 samples corresponding to different values of $\alpha$. The numerical results for each trajectory are shown in Fig.~\ref{fig:DITCP}b. 
The data display $\Delta_D$-dependence (not $2\Delta_0$ scale) of the Chern number. 
Furthermore, Fig.~\ref{fig:DITCP}c shows the sum of the Chern number between $C_{\rm outer}$ and $C_{\rm inner}$ trajectories. The result clearly captures the presence of a non-trivial pump between $10E_R\lesssim V_D \lesssim 30E_R$, quantified by the imperfect cancellation between the Chern numbers for $C_{\rm outer}$ and $C_{\rm inner}$ trajectories.
Note that the experimental trajectory surely yields the disorder-induced pumping;
however, this does not necessarily mean that this pump is topological.
It could be topological if we adjust appropriately the lattice parameters including the phase connecting the outer and inner trajectories.


\subsection*{Discussion}
Our unique experimental platform provides us with interesting opportunities for studying a wide range of topological quantum phenomena with disorder. 
By introducing a disorder lattice with $\lambda_D \simeq 1217$~nm,
it is possible to add a quasi-periodic disorder where $d_D/d$ approximates the golden ratio, 
which is often studied in the Aubry-Andr\'e model~\cite{AAmodel}, or to add a genuinely random (non quasi-periodic) disorder by using a speckle pattern.
Moreover, in this study we connect two non-trivial pump trajectories to create a trivial trajectory
in the clean limit and then add disorder to observe disorder-induced pumping.
This method presented here allows the realization of disorder-induced pumping, and will give a suggestion or guideline for the future study of a wide range of disorder-induced topological phases.
Our highly controllable optical lattice system loaded with ultracold fermions can straightforwardly incorporate interatomic on-site interaction with reasonably large strengths compared with tunneling energy.
Consequently, we can study the effect of the disorder on Thouless pump in
a strongly correlated regime \cite{Nakagawa, Stenzel}, especially in the regime of many-body localization~\cite{Huse,Nandkishore} which may protect topological phenomena up to higher excited states. The pumping can be extended to higher dimensions~\cite{Lohse2018, Zilberberg2018, Price2015}. For example, the two-dimensional extension of our experiment is directly related to disordered quantum Hall system \cite{Ippoliti2020}.
The introduced disorder can also be dynamically controlled, 
enabling us to study an anomalous Floquet Anderson topological phase \cite{Titum} only possible for a Floquet system, revealed by nonadiabatic pumping.
Our experimental setup can also allow us to study the effect of nonadiabaticity~\cite{Citro2018}.
Moreover, the effect of disorder on the higher-order topological phenomena~\cite{Benalcazar61,Xie,SQShen} is an interesting new research direction~\cite{Araki}. 

\section{Methods}    
\subsection{Preparation of a degenerate Fermi gas of $^{171}$Yb}
Since $^{171}$Yb atoms have a very small $s$-wave scattering length of -0.15~nm~\cite{Kitagawa08},
they can be regarded to be non-interacting. Therefore,
we use sympathetic evaporative cooling with $^{173}$Yb atoms to obtain 
the degenerate Fermi gas of $^{171}$Yb~\cite{Taie10}.

The number of atoms for each spin is typically 3500.
A typical atom temperature before lattice loading is $T/T_F \sim 0.24 $ for the measurements in Figs.~2-4 and $\sim0.19$ for Fig.~5 at the end of the evaporation with the trap frequencies of the far-off resonant optical trap of
$(\omega_x', \omega_y', \omega_z)/2\pi=(119, 25 ,153)$~Hz, 
where the $x'$- and $y'$-axes are tilted from the lattice axes ($x$ and $y$) by $45^\circ$. 
The estimated sizes of our atomic clouds (full width at half maximum, FWHM) are about $4.4~\mu\mathrm{m}$, $21~\mu\mathrm{m}$, and $3.4~\mu\mathrm{m}$ for the $x'$, $y'$, and $z$ directions, respectively.
The system size along the pumping direction ($x$) is about $6.3~\mu\mathrm{m}$, corresponding to about 12 unit cells around the center of the atomic clouds.
The whole atomic system consists of about $60 \times 6$ 1D lattices along the $y$- and $z$-axes, and totally about 360 lattices.

\subsection{Setup for the cRM superlattice and quasi-periodic disorder lattice}
An incommensurate optical lattice potential created by a retro-reflected laser beam (wavelength at $\lambda_D$) and tilted by 45 degrees to the pumping direction
is superimposed on a 2D array of one-dimensional cRM superlattice.
This superimposed optical lattice (disorder lattice) 
creates a periodic potential whose lattice constant $d_D=\lambda_D/\sqrt{2}$ in the pumping direction of the cRM, acting as a quasi-periodic disorder. 
In our setup, the wavelength $\lambda_D$ can range from 776~nm to 820~nm realized by a Ti:Sapphire laser,
and the relative incommensurate lattice constant $d_D/d$ ranges from 1.03 to 1.09, respectively.
Because the ratio of this wavelength is relatively close to 1, this quasi-periodic disorder does not take the same structure in the trap region. Also, because this lattice is tilted by 45 degrees, the phase of the quasi-periodic disorder shifts by 3.05, 2.96, and 2.88~rad for $\lambda_D=$ 776, 798, and 820~nm, respectively, in the adjacent array (Fig.~\ref{fig:RM}a).

\subsection{Pumping speed and adiabadicity}
An essential requirement of the topological Thouless pump is adiabaticity.
This is necessary to avoid non-adiabatic Landau-Zener transitions to the higher band during the pumping process, at least in the clean limit.
The pumping times per cycle take 130~ms on average for the normal charge pumping in Figs.~\ref{fig:dis_dep}-\ref{fig:WL_dep} and 460~ms for the disorder-induced pumping in Fig.~\ref{fig:DITCP}, 
in particular 100~ms for outer and inner trajectories and 130~ms for the connecting region (two times), respectively. 
These pumping times are sufficiently longer than the Landau-Zener-transition time scales, which are estimated to be $\sim$1~ms based on the maximum and minimum bandgaps during the pumping cycles~\cite{Nakajima}. 

\section*{Additional information}
Correspondence and requests for materials should be addressed to S. Nakajima.

\section*{Acknowledgements}
We thank H.~Aoki, Y.~Hatsugai, and K.~Imura for valuable discussions
and A.~Sawada for experimental assistance.
This work was supported by the Grant-in-Aid for Scientific Research of JSPS (Nos.~JP25220711, JP26247064, JP16H00990,
JP16H01053, JP17H06138, JP18H05405, JP18H05228, and JP18K13480), 
the Impulsing Paradigm Change through Disruptive Technologies (ImPACT) program, 
JST CREST (No.~JPMJCR1673), and MEXT Quantum Leap Flagship Program (MEXT Q-LEAP) Grant No.~JPMXS0118069021.
Y.~K. acknowledges the support of the Grant-in-Aid for JSPS Fellows (No.17J00486).
P.~M. is supported by the Japan Science and Technology Agency (JST) of the Ministry of Education, Culture, Sports, Science and Technology (MEXT), 
JST CREST Grant. No. JPMJCR19T2 and 
by the (MEXT)-Supported Program for the Strategic Research Foundation at Private Universities “Topological Science” (Grant No. S1511006)
and by JSPS Grant-in-Aid for Early-Career Scientists (Grant No. 20K14375).

\section*{Author contributions}
S.~N. and N.~T. contributed equally to this work.
S.~N., N.~T. and K.~S. carried out experiments and the data analysis.
Y.~K. and P.~M. carried out the theoretical calculation.
Y.~T. conducted the whole experiment. All the authors contributed to the writing of the manuscript.

\section*{Competing interests}
The authors declare no competing interests.

\section*{References}
\providecommand{\noopsort}[1]{}\providecommand{\singleletter}[1]{#1}%


\onecolumngrid
\clearpage
\begin{center}
\noindent\textbf{Supplementary Information for:}
\\\bigskip
\noindent\textbf{\large{Competition and interplay between topology and quasi-periodic disorder in Thouless pumping of ultracold atoms}}
\\\bigskip
Shuta~Nakajima$^{1,2}$, Nobuyuki~Takei$^{1}$, Keita~Sakuma$^{1}$, Yoshihito~Kuno$^{1,3}$, Pasquale~Marra$^{4,5}$, and Yoshiro Takahashi$^{1}$ \\
\vspace{0.1cm}
\small{$^1$ \emph{Department of Physics, Graduate School of Science, Kyoto University, Kyoto 606-8502, Japan}}\\
\small{$^2$ \emph{The Hakubi Center for Advanced Research, Kyoto University, Kyoto 606-8502, Japan}}\\
\small{$^3$ \emph{Department of Physics, University of Tsukuba, Tsukuba, Ibaraki 305-8571, Japan}}\\
\small{$^4$ \emph{Graduate School of Mathematical Sciences, The University of Tokyo, Komaba, Tokyo, 153-8914, Japan}}\\
\small{$^5$ \emph{Department of Physics, and Research and Education Center for Natural Sciences, Keio University, Hiyoshi, Kanagawa, 223-8521, Japan}}
\end{center}
\bigskip
\bigskip
\twocolumngrid

\newcommand{\figureref}[1]{Figure \ref{#1}}
\renewcommand{\thefigure}{S\the\numexpr\arabic{figure}-10\relax}
\setcounter{figure}{10}
\renewcommand{\theequation}{S\the\numexpr\arabic{equation}-10\relax}
\setcounter{equation}{10}
\renewcommand{\thesection}{S\arabic{section}}
\setcounter{section}{0}
\renewcommand{\bibnumfmt}[1]{[S#1]}
\renewcommand{\citenumfont}[1]{S#1}

\section{Approximated tight-binding model}    
In order to estimate the qualitative properties of our experimental system modeled by the cRM model, 
we consider an approximated tight-binding picture.
Our experimental system is based on the following dynamical superlattice potential in a clean limit:
\begin{eqnarray}\label{SL}
V_\mathrm{clean}(x,t) = -V_{S}\cos^{2} \biggl(\frac{2\pi x}{d}\biggl)\!-\! V_{L}\cos^{2} \biggl(\frac{\pi x}{d} \! -\phi(t) \hspace{-2pt} \biggl).\nonumber \\
\end{eqnarray}
Hereafter, since in our experiments only the first band is almost occupied, and the band gap between the second and third bands are fairly large, 
we can focus on only the first and second bands created by our experimental superlattice potential. 
Then, we can describe the system with an effective tight-binding model, namely, the tRM model in a clean limit, which reads 
\begin{eqnarray}
{\hat H}_{\rm RM}=\sum_{i}\biggl[&-&(J+\delta)\hat{a}^{\dagger}_{i}{\hat b}_{i}-(J-\delta)\hat{a}^{\dagger}_{i}{\hat b}_{i+1}+\mbox{h.c.} \nonumber \\
&+&\Delta (\hat{a}^{\dagger}_{i}{\hat a}_{i}-\hat{b}^{\dagger}_{i}{\hat b}_{i})\biggl].
\label{RM}
\end{eqnarray}
In the topological charge pumping (TCP), $\delta$ and $\Delta$ are adiabatic dynamical parameters, 
which can be approximately determined by the band structure induced by the superlattice potential $V_\mathrm{clean}(x,t)$. 
Here, we approximate these dynamical sequence for $\delta$ and $\Delta$ as a circular trajectory, $\delta=\delta_{0} \sin\theta$, $\Delta(t)=\Delta_{0} \cos\theta$, 
where $\theta=2\phi(t)$.
Although the actual dynamical sequence for $\delta$ and $\Delta$ in our experimental system is subtly different from the circular one, 
this approximation facilitates the ideal theoretical treatment of the TCP phenomena, especially for calculating the Chern number, 
and can gives insight for qualitative properties of the TCP described by the cRM in our experiment. 
The physical values of $J$, $\delta_{0}$ and $\Delta_{0}$ can be directly determined by the band structure of the superlattice potential $V_\mathrm{clean}(x,t)$. 
The values of $J$ and $\delta_{0}$ are determined by the band width in the double-well lattice case ($\theta=\pi/2$, $3\pi/2$), 
while the value of $\Delta_{0}$ by the band gap in the staggered lattice case ($\theta=0$, $\pi$) as discussed in the previous study \cite{Nakajima}. 
For the band structure determined by our experimental setup, the tRM parameters are approximately determined. 
The tRM model is expected to capture the qualitative behavior for our experimental system, 
since in the clean limit, the circular parameter sequence for $\delta$ and $\Delta$ is topologically connected to our actual experimental sequence without gap closing. 
Regarding the basic topological aspect of the tRM model, for $\Delta=0$ (at inversion symmetric point, $\theta=\pi/2$, $3\pi/2$), 
the model reduces to the celebrated Su-Schrieffer-Heeger (SSH) model, classified in the Altland-Zirnbauer class BDI of the periodic classification of topological states\cite{Schnyder,Kitaev}. 
The SSH model exhibits a topological phase and zero energy topological edge modes \cite{Asboth}. 
However, a finite $\Delta$ breaks the chiral symmetry without gap closing, i.e., 
${\hat \sigma}_{z}{\hat H}_{\rm RM}(k){\hat \sigma}_{z}\neq -{\hat H}_{\rm RM}(k)$ 
where ${\hat H}_{\rm RM}(k)$ is the bulk momentum Hamiltonian of Eq.~(\ref{RM}). 
Accordingly, the tRM model is not necessarily in the BDI class.

\begin{figure*}[t]
\centering
\includegraphics[width=10cm]{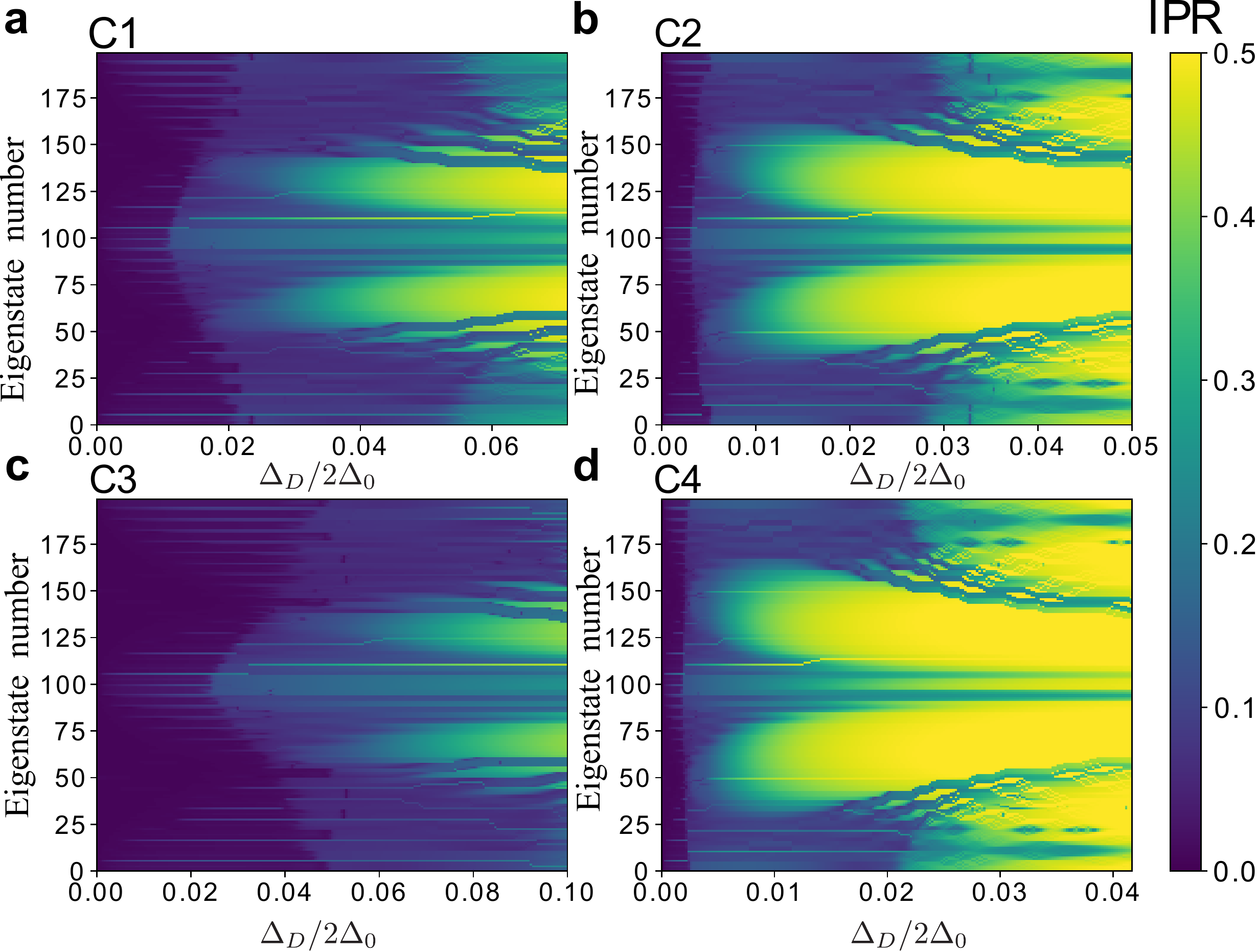}
\caption{{\bf IPR calculation for a quasi-periodic disordered SSH model.}
{\bf a}, $C_1$: ($V_{L}$,$V_{S})=(20,14)E_R$, ($J$,$\delta_{0}$,$\Delta_{0})=(0.861,0.851,6.45)E_R$. 
{\bf b}, $C_2$: ($V_{L}$,$V_{S})=(30,20)E_R$, ($J$,$\delta_{0}$,$\Delta_{0})=(0.860, 0.857, 8.54)E_R$. 
{\bf c}, $C_3$: ($V_{L}$,$V_{S})=(15,10)E_R$, ($J$,$\delta_{0}$,$\Delta_{0})=(0.914, 0.898, 4.86)E_R$. 
{\bf d}, $C_4$: ($V_{L}$,$V_{S})=(36,24)E_R$, ($J$,$\delta_{0}$,$\Delta_{0})=(0.827, 0.825, 9.605)E_R$. }
\label{disorder_SSH}
\end{figure*}

In our experiment, a quasi-periodic disorder lattice created by an additional laser with wavelength $\lambda_D$ (= 798~nm) applied by tilting 45$^\circ$ for the one-dimensional superlattice axis, is given by 
\begin{eqnarray}\label{DL}
V_{\rm Dis}(x)=-V_{D}\cos^{2}\biggl(\frac{\pi x}{d_{D}}+\frac{\alpha}{2} \biggl),
\end{eqnarray}
(see Fig.~1a of the main text).
This quasi-periodic disorder potential $V_{\rm Dis}(x)$ can be effectively implemented in the tRM model as an on-site potential given by
\begin{eqnarray}\label{VDis}
\hat{V}'_{\rm Dis}=&&\frac{\Delta_D}{2}\sin [2\pi \beta (2i)+\alpha]{\hat a}^{\dagger}_{i}{\hat a}_{i} \nonumber \\
&&+\frac{\Delta_D}{2}\sin [2\pi \beta (2i-1)+\alpha]{\hat b}^{\dagger}_{i}{\hat b}_{i},
\end{eqnarray}
where $\beta=d/(2 d_D)$.
The potential amplitude $\Delta_D$ in $\hat{V}'_{\rm Dis}$ does not strictly correspond to the value of the continuous system potential $V_D$, but gives a good approximation.
The quasi-periodic disorder potential $\hat{V}'_{\rm Dis}$ is diagonal and breaks the chiral symmetry even for $\theta=\pi/2$ or $2\pi/3$ SSH case. 
In general, this type of the disorder strongly affects the topological system and leads to the breakdown of the topological phase, 
compared to a chiral symmetric disorder employed in a recent experiment \cite{Meier}. 
On the other hand, in this tight-binding picture we ignore modulated effects of $V_{\rm Dis}(x)$ to the tRM parameters, $J$, $\delta_{0}$ and $\Delta_0$. 
Thus, we consider ${\hat H}_{RM}+\hat{V}'_{\rm Dis}$ as an approximated tight-binding model for our experimental system.

For the sake of simplicity, we consider the effect of quasi-periodic disorder only on the on-site potential and ignore the effects of the trapping potential.
However, the resulting model is nevertheless able to capture the essence of the physical behavior of the cRM model and of the experimental system. In particular, for small $V_D$, the tRM model reasonably approximates the cRM model.
However, for large $V_D$, deviations between the tRM model and cRM models become noticeable.
Since $\hat{V}'_{\rm Dis}$ is diagonal, the total tight-binding model has no chiral symmetry even for inversion symmetric point, 
therefore $\hat{V}'_{\rm Dis}$ is expected to have strong influence to the TCP.  

\section{Localization transition in a quasi-periodic disordered SSH model}
To estimate the Anderson localization point $V_{AL}$ in our experimental system, 
we calculate effects of quasi-periodic disorder created in a disordered SSH model, corresponding to the $\Delta=0$ case in the tRM model of Eq.~(\ref{RM}) including the $\hat{V}'_{\rm Dis}$ of Eq.~(\ref{VDis}) as an effect of the disordered lattice potential.
Here, we consider a quasi-periodic disorder lattice with a wavelength of $798$~nm.
In general, since we consider quasi-periodic disorder, 
we expect an Anderson localization transition at a finite disorder strength even in the one dimensional system.
In conventional  Anderson localization, single atoms are localized at each single lattice site.
To estimate the localization transition in our considering system, we employ the inverse participation ratio (IPR).
The IPR for each eigenstate is defined as 
\begin{eqnarray}
({\rm IPR})_{\ell}=\sum_{j}|\langle j|\psi_{\ell}\rangle|^{4},
\end{eqnarray} 
where $|j\rangle$ is a $j$-site localized single particle state and $|\psi_{\ell}\rangle$ is $\ell$-th eigenstate in the SSH model with quasi-periodic disorder potential.
Large value of the IPR is a signature of the localized tendency of the single particle wave function. 
We perform the IPR calculation for four different parameter sets of ($J$,$\delta$) 
corresponding to those in the four different pump trajectories in our experiments.
The results are shown in Fig.~\ref{disorder_SSH}, where we plot the IPR as a function of $\Delta_{D}$ and of the eigenstate number in $L=200$ lattice site system with periodic boundary conditions. 
The sets $C_1$ and $C_2$ correspond to the parameter sets A and B in Fig.~2a in the main text, and the set $C_3$ and $C_4$ correspond to the trajectories $C_\mathrm{inner}$ and $C_\mathrm{outer}$ in Fig.~5b, respectively. 
The order of the eigenstate numbers corresponds to the eigenenergy in ascending order. 
Due to the quasi-periodic disorder, all cases in Fig.~\ref{disorder_SSH} exhibit mobility edges, 
i.e., different eigenstates localize for different disorder strengths. 
Here we take the condition $({\rm IPR})_{\ell}\gtrsim 0.1$ as a localization criterion, while the transition point is determined by the case where all eigenstates fulfill this condition. 
For all the cases in Fig.~\ref{disorder_SSH}, all the eigenstates already tend to be localized for $\Delta_{D}/2\Delta_0 <0.04$. 
We define the Anderson localization transition point $V_{AL}$ in the main text, using eigenstates with lower eigenenergies.
These states are localized at $\Delta_D/2\Delta_0 \sim 0.02$ and $\Delta_D/2\Delta_0 \sim 0.004$ for the pump trajectories $C_1$ and $C_2$ (or the set A and B in the main text), respectively.
These values correspond to $V_{AL}^A\sim 0.3E_R$ and $V_{AL}^B\sim 0.07E_R$.
Even though from the numerical calculation our experimental system (cRM model) is also expected to be localized around the SSH parameters points, 
this fact does not mean gap closing directly. 
Even when the system exhibits localization and fulfills our criteria,
the pumping does not necessarily break down, as shown numerically in Ref.~\cite{Wauters}.

\section{Chern number calculation from a dimensional extended tight-binding model} 
The tRM model can be mapped into the Harper--Hofstadter--Hatsugai (HHH) model \cite{Hatsugai} through a dimensional extension \cite{Kraus,Lohse}.
By regarding the pumping parameter $\theta$ in ${\hat H}_{\rm RM}$ of Eq.~(\ref{RM}) as $y$-direction momentum $k_{y}$ and applying the Fourier transformation for $k_{y}$, 
we can obtain the HHH model from the tRM model:
\begin{eqnarray}
\hat{H}_{2D}&=&\sum_{m,n}\biggl[-J\hat{c}^{\dagger}_{m+1,n}\hat{c}_{m,n}
                 -\frac{\Delta_{0}}{2}e^{-i\pi(m-1/2)}\hat{c}^{\dagger}_{m,n+1}\hat{c}_{m,n}\nonumber\\
                &&-\frac{\delta_{0}}{2}e^{i\pi m}\hat{c}^{\dagger}_{m+1,n+1}\hat{c}_{m,n}
                 -\frac{\delta_{0}}{2}e^{-i\pi m}\hat{c}^{\dagger}_{m+1,n}\hat{c}_{m,n+1}\nonumber\\
                 &&+\mbox{h.c.}\biggr].
\label{HHH}
\end{eqnarray}
Here $\hat{c}^{(\dagger)}_{m,n}$ is the annihilation (creation) operator at a two dimensional lattice site $(m,n)$, 
where $m=1,2,\cdots, L_{x}$ and $n=1,2,\cdots, L_{y}$, i.e., $L_{x}\times L_{y}$ lattice system.
The HHH model is known to have two topological bands with a non-zero Chern number \cite{Hatsugai,Kraus}. 
This is a peculiar property compared to a conventional quantum Hall system on lattice described by the Hofstadter model \cite{Hofstadter},
since the Hofstadter model does not exhibit two separated bands with non-trivial topology. 
The two topological bands with a non-zero Chern number in the HHH model is the origin of the TCP in the tRM model. 
The value of the Chern number corresponds to the total pumped current per one pumping cycle in the tRM model \cite{Asboth}.
For the HHH model, we can further implement the effect of the quasi-periodic disorder $\hat{V}'_{dis}$ of the tRM model 
as $\hat{V}''_{\rm Dis}=\frac{\Delta_D}{2}\sin (2\pi \beta m+\alpha){\hat c}^{\dagger}_{n,m}{\hat c}_{n,m}$. 
Here, it should be noted that $\hat{V}''_{\rm Dis}$ depends only on the $x$-component site $m$ and is uniform along $y$-direction. 

\begin{figure}[t]
\centering
\includegraphics[width=8cm]{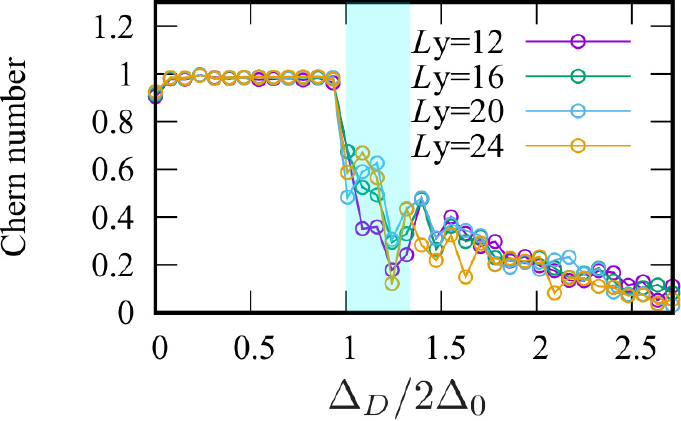}
\caption{{\bf Numerical calculation of Chern number in the presence of quasi-periodic disorder and its dependence on $L_y$.}
$L_{x}$ is fixed to $40$.
We average over 20 different values of $\alpha$, the tRM parameters are set to the $C_{1}$ case.
The blue shaded regime represents a theoretical expected phase transition regime, where the TCP is suddenly suppressed.}
\label{FSE_CN}
\end{figure}

In our experimental system, atoms almost occupy the lowest band and the occupation of atom in higher bands is fairly suppressed. 
We assume that the lower band of the tRM model is fully-occupied approximately. 
Accordingly, in the numerical calculation of the HHH model the half-filling case is considered. 
Then, to make comparison with the experimental results shown in the main text, 
we set the HHH model parameters ($J$, $\delta_0$, $\Delta_0$) to four experimental parameter sets: 
(I) ($V_{L}$,$V_{S})=(20,14)E_R$, ($J$,$\delta_{0}$,$\Delta_{0})=(0.861,0.851,6.45)E_R$, 
(II) ($V_{L}$,$V_{S})=(30,20)E_R$, ($J$,$\delta_{0}$,$\Delta_{0})=(0.860, 0.857, 8.54)E_R$,
(III) ($V_{L}$,$V_{S})=(15,10)E_R$, ($J$,$\delta_{0}$,$\Delta_{0})=(0.914, 0.898, 4.86)E_R$, and
(IV) ($V_{L}$,$V_{S})=(36,24)E_R$, ($J$,$\delta_{0}$,$\Delta_{0})=(0.827, 0.825, 9.605)E_R$.
These approximated circular pumping trajectories determined by the above four parameter sets are denoted by $C_{1}$, $C_{2}$, $C_{3}$ and $C_{4}$.  
These parameter sets ($J$, $\delta_0$, $\Delta_0$) are determined by comparing to the energy spectra of the cRM model as in the previous paper \cite{Nakajima}. 
The $C_1$ and $C_2$ trajectories are the approximated version of the set A and B in the experimental pumping sequences in Fig.~2a in the main text, respectively. 
The $C_3$ and $C_4$ trajectories correspond to the experimental pumping sequences $C_\mathrm{inner}$ and $C_\mathrm{outer}$ in Fig.~5b, respectively.
In what follows, We numerically calculated the Chern number for $C_1$, $C_3$ and $C_4$ trajectories and use $L_x=40$ (20 unit cells). 
This $x$-direction system size is close to our experimental system.

We calculate the Chern number for the model of $\hat{H}_{2D}+\hat{V}''_{\rm Dis}$ in the following manner. 
Since discrete translational invariance is broken due to the term $\hat{V}''_{\rm Dis}$, 
we employ a calculation method to obtain the Chern number from the real space Hamiltonian, 
namely, the so-called coupling matrix method \cite{YFZhang,Castro,Sriluckshmy,Kuno}. 
In this method, we impose twisted periodic boundary conditions with two twisted phase $\theta_{x}$ and $\theta_{y}$ for each spatial direction. 
Their twisted phases play a role of the momentum $k_{x}$ and $k_{y}$ (corresponding to $\theta$) in the TKNN formula \cite{Niu85}. 
Since this twisted phase boundary conditions can be employed even for a system without translational invariant such as a disordered system, 
we can calculate the Chern number for the model of $\hat{H}_{2D}+\hat{V}''_{\rm Dis}$.

\begin{figure*}[t]
\centering
\includegraphics[width=16cm]{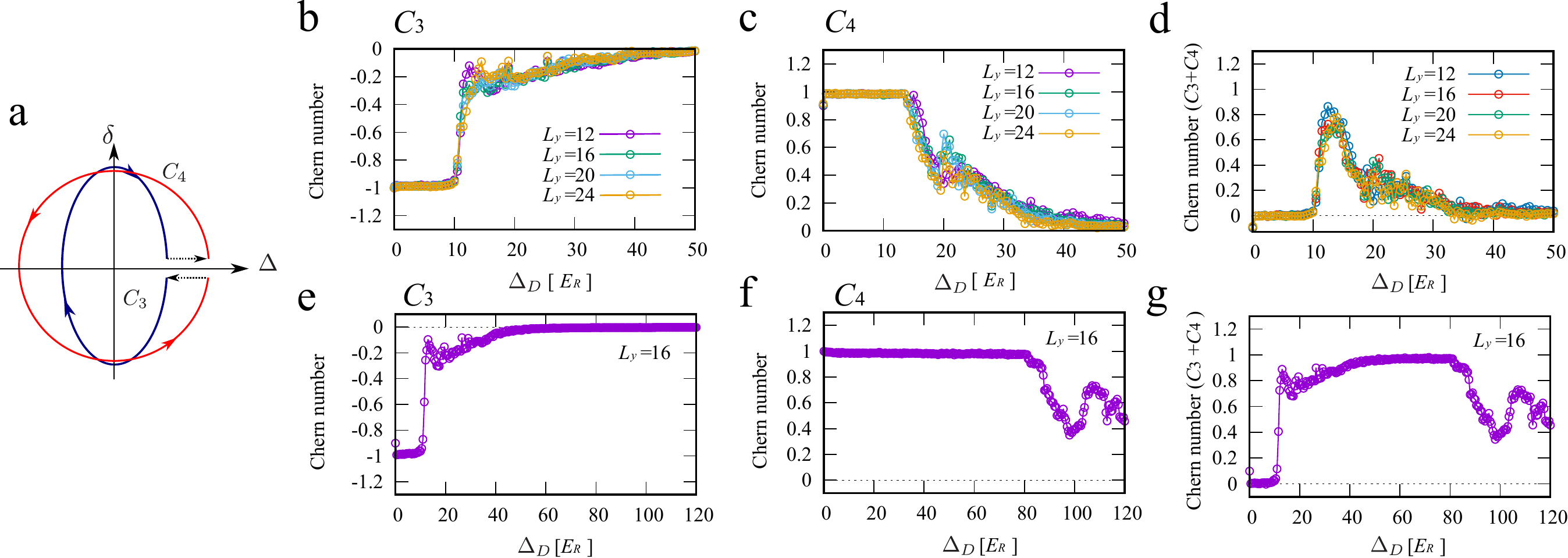}
\caption{
{\bf Numerical calculation of Chern number in the disorder-induced charge pumping (DICP).}
{\bf a}, Pumping trajectory. This is created by combining the two circular trajectories $C_{3}$ and $C_{4}$ with different circular direction. 
The two trajectories are connected by changing the staggered potential $\Delta_{0}$ with fixed $\delta=0$. 
The total pumping trajectory does not wrap the origin corresponding to the gap closing point of the clean RM model.
Without the quasi-periodic disorder lattice, the total charge pumping along this trajectory is zero. 
{\bf b}, Chern number behavior for $C_{3}$ trajectory and
{\bf c}, for $C_{4}$ trajectory as a function of $\Delta_D$.
The TCP in $C_{4}$ case is more robust than in $C_{3}$ trajectory.
{\bf d}, The difference of the Chern number between $C_{3}$ and $C_{4}$ trajectories as a function of $\Delta_D$.
The numerical result indicates possibility to exhibit a disorder-induced pumping for $10E_R \lesssim \Delta_D \lesssim 30E_R$. $L_y$-system size dependence is fairly small.
For the case that the inner ($C_{3}$) and outer ($C_{4}$) trajectories are farther apart:
{\bf e}, Chern number behavior for $C_{3}$ trajectory and
{\bf f}, for $C_{4}$ trajectory as a function of $\Delta_D$.
The TCP in $C_{4}$ case is much more robust than in $C_{3}$ trajectory.
{\bf g}, the difference of the Chern number between $C_{3}$ and $C_{4}$ trajectories as a function of $\Delta_D$.
The numerical result indicates the possibility to realize a topological DICP for $40E_R \lesssim \Delta_D \lesssim 80E_R$, where a clear quantized plateau appears.}
\label{C3-C4}
\end{figure*}

Figure~\ref{FSE_CN} displays the calculated Chern number in the presence of quasi-periodic disorder and its dependence of the artificial $y$-spatial system size $L_{y}$ with fixed $L_{x}=40$ (20 unit cells system from the view point of the tRM model).  
The result indicates that the system-size dependence is fairly small, 
and the topological phase transition point does not depend much on the value of $L_{y}$. 
Therefore, the system size used in the numerical calculation is large enough to capture the behavior of our experimental system.
$L_y=24$ data in Fig.~\ref{FSE_CN} is displayed in Fig.~2b in the main text.

Next, we employ this numerical method to study the disorder-induced charge pump (DICP). 
The schematic pumping trajectory plotted in the $\delta-\Delta$ plane is shown in Fig.~\ref{C3-C4}a. 
This trajectory is a combination of circular $C_3$ and $C_{4}$ trajectories.
Our experiment is conducted with a trajectory similar to this schematic trajectory.
In particular, the experimental trajectory can be connected by continuous deformation, and it is therefore topologically equivalent, to the trajectory shown in Fig.~\ref{C3-C4}a.
The DICP trajectory does not wrap the origin $(\Delta_{0}, \delta_{0})=(0,0)$ corresponding to the gap closing point in the tRM model in the clean limit.
In this sense, in the clean limit, the tight-binding RM model of Eq.~(\ref{RM}) exhibits no TCP, i.e., total pumped current is zero. 
However, once the quasi-periodic disorder $\hat{V}'_{\rm Dis}$ is switched on, 
the disordered model has possibility to exhibit a non-trivial pumped current that could not occur at the clean limit. 
This corresponds to the fact that the disordered HHH model also has possibility to exhibit a topological non-trivial phase with a non-vanishing Chern number, $C_{N}\neq 0$.

To study the DICP, we split the DICP trajectory into two circles, $C_3$ and $C_{4}$. 
For each circular trajectory, we calculate the Chern number by using the disordered HHH model. 
Here, various system sizes for $L_{y}$ are calculated to check the finite size effects. 
Moreover, for all data we average over different samples  corresponding to different values of $\alpha$ (60 samples). 
The numerical results for each trajectory are shown in Fig.~\ref{C3-C4}b and c. 
The data display $\Delta_D$-dependence (not $2\Delta_{0}$ scale) of the Chern number. 
Furthermore, Fig.~\ref{C3-C4}d shows the sum of the Chern number between $C_{3}$ and $C_{4}$ trajectories. 
The result clearly captures the presence of a non-trivial pump between $10E_{\rm R}\lesssim \Delta_D\lesssim 30E_{\rm R}$, 
quantified by the imperfect cancellation between the Chern numbers for $C_{3}$ and $C_{4}$ trajectories. 
All results in Fig.~\ref{C3-C4}b-d indicate that the finite system size ($L_{y}$) dependence is fairly small. 
$L_y=24$ data in Fig.~\ref{C3-C4} are used in Fig.~5b and c in the main text.

In addition, we show that this type of pump protocol in Fig.~\ref{C3-C4}a theoretically exhibits a topological DICP.  
Here, we consider that the inner and outer trajectories are farther apart compared with the experimental ones, i.e., we set the inner circle (C3) with $(J, \delta_0, \Delta_0)=(1,0.9,5)E_R$ and the outer circle (C4) with $(J,\delta_0, \Delta_0)=(6,5,45)E_R$. 
The numerical results are shown in Fig.~\ref{C3-C4}e-g. 
As seen in Fig.\ref{C3-C4}g, there exists a quantized plateau within the errors or fluctuations due to numerical instability inevitable for disordered systems for moderate disorder strength.

\begin{figure*}[t]
\centering
\includegraphics[width=16cm]{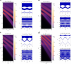}
\caption{
{\bf Bulk density of states (DoS) as a function of $V_\mathrm{D}$
and  band structure in the clean and in the strong disorder regimes,
for different values of the lattice depths $V_\mathrm{L}$ and $V_\mathrm{S}$.}
{\bf a}, For $V_\mathrm{L},V_\mathrm{S}=5, 3.5$ we can clearly see from the DoS that the global gap closes and reopens at $V_\mathrm{D}\approx 5$.
{\bf b}, Also for $V_\mathrm{L},V_\mathrm{S}=7.5, 5$ we can clearly see that the global gap closes and reopens at $V_\mathrm{D}\approx 5$.
{\bf c}, For $V_\mathrm{L},V_\mathrm{S}=10, 7$ the gap closes and a very small gap reopens at $V_\mathrm{D}\approx 7$.
{\bf d}, For $V_\mathrm{L},V_\mathrm{S}=20, 14$ which is the same as Fig.~3d of the main text, the gap closes at $V_\mathrm{D}\approx 5$ but does not reopen.
We consider a system with total length equal to $198 d$ and with $\alpha=0$. The DoS is averaged over the phase $\phi$.}
\label{fig:CM}
\end{figure*}
\begin{figure}
\centering
\includegraphics[width=8.5cm]{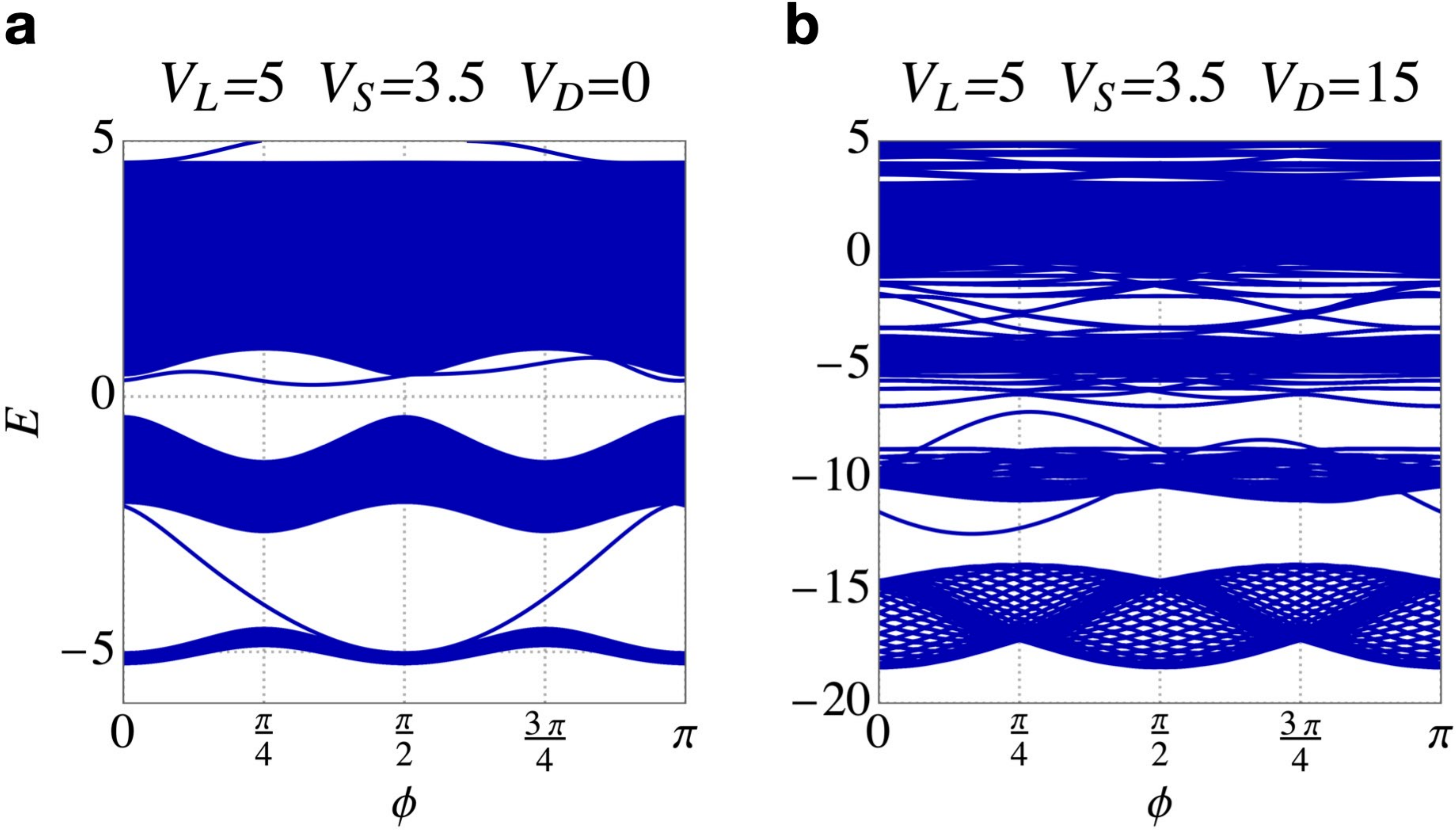}
\caption{
{\bf Band structure for a system with open boundary conditions with $V_\mathrm{L},V_\mathrm{S}=5, 3.5$.}
{\bf a}, In the clean limit $V_\mathrm{D}=0$ there is a non-trivial edge state connecting the lowest energy band (at filling $j=1$) with the first excited band.
{\bf b}, In the strong disorder regime $V_\mathrm{D}=15$ there are no edge states connecting the lowest energy band with the first excited band.
}
\label{fig:CM2}
\end{figure}

\section{Continuous model}
To obtain the band structure in the continuous system, we use a continuous model with a lattice potential given by
\begin{eqnarray}\label{latt}
V(x,t)= 
&-& V_\mathrm{S}\cos^2\left(\frac{2\pi x}{d}\right) 
-V_\mathrm{L}\cos^2\left(\frac{\pi x}{d}-\phi(t) \right)\nonumber \\
&-&V_\mathrm{D}\cos^2\left(\frac{\pi x}{d_\mathrm{D}} +\frac{\alpha}{2} \right)
,
\end{eqnarray}
and we solve the time-independent Schr{\"o}dinger equation
\begin{equation}
\left[
-\frac{\hbar^2\partial^2_x}{2m}
+V(x,\phi)
\right]
\Psi(x,t)=
E \Psi(x,t),
\end{equation}
using space discretization.
Using the recoil energy as the energy unit and the lattice constant of the long lattice 
as the length unit one has
$E_R={h^2}/{(8m)^2}=1$.

Since $d_\mathrm{D}/d$ is an irrational number, the quasi-periodic disorder lattice is incommensurate with respect to the short and the long lattice.
As a consequence, the translational symmetry of the lattice is broken, and thus the unit cell of superlattice is not well-defined.
Hence, one cannot meaningfully define the periodic boundary conditions for the system.
Therefore, we will approach the irrational value $d_\mathrm{D}/d={3}/{2\sqrt{2}}$ by taking a succession of rational approximations of the irrational number, obtained in terms of continued fraction representation~\cite{continued_fractions_hardy}.
Every irrational number $R\in\mathbb{R}-\mathbb{Q}$  can be written uniquely as an infinite continued fraction as
\begin{equation}
R=[N_0; N_1, N_2, \,\ldots ] =N_0+\cfrac{1}{N_1 + \cfrac{1}{N_2 +   {\dots}}}
\end{equation}
with $N_i\in\mathbb{Z}$ integers.
The successive approximations $R_n$ are obtained by truncating the continued fraction representation $[N_0; N_1, N_2, \,\ldots, N_n]$.
Since $R_n$ are rational numbers one can write as
\begin{equation}
R_n=[N_0; N_1, N_2, \,\ldots N_n ] =\frac{P_n}{Q_n}\approx R,
\end{equation}
where $P_n, Q_n$ are coprimes.
These represent successive rational approximations of the irrational number $R$.

In our specific case one has 
\begin{eqnarray}
\frac{d_\mathrm{D}}{d}&=&\frac{3}{2\sqrt{2}}
\approx
\frac{17}{16},\frac{35}{33},\frac{577}{544},
\ldots \nonumber \\
&=&
1.0625,
1.060606060\ldots,
1.060661764\ldots,
\ldots \nonumber \\
\end{eqnarray}
For $V_\mathrm{D}=0$ the unit cell of the superlattice has length $d$. 
For $V_\mathrm{D}>0$ and $R=d/d_\mathrm{D}$ being a rational number $R=P/Q$ with $P,Q$ coprimes, the total unit cell of the superlattice has length $Q d$ as one can see by direct substituting $x\to x+Q$ Eq.~(\ref{latt}).
The energy levels are then obtained by solving the  Schr{\"o}dinger equation with periodic boundary conditions over a system of length $L=N_\mathrm{c} Q d$ where $N_\mathrm{c}$ is the total number of unit cells.
We take $P/Q=35/33$ and a total length of $198 d$.

Figure~\ref{fig:CM} shows the bulk
density of states (DoS) as a function of $V_\mathrm{D}$
and the band structure in the clean and in the strong disorder regimes,
for different values of the lattice depths $V_\mathrm{L}$ and $V_\mathrm{S}$.
The DoS and the band structure are calculated using closed (periodic) boundary conditions. 
In the shallower lattice cases (a and b) we can clearly see from the DoS that the global gap closes and reopens at $V_\mathrm{D}\approx 5$.
For the case (c) , a very small gap closes and reopens at $V_\mathrm{D}\approx 7$.
In the case (d) , which is the same as Fig.~3d of the main text, 
the numerical calculations show the gap closing at $V_\mathrm{D}\approx 5$ but do not show such re-opening of the gap.

Figure~\ref{fig:CM2} shows the band structure calculated in the case of open boundary conditions in the clean $V_\mathrm{D}=0$ and strong $V_\mathrm{D}=15$ disorder regimes, for the parameter set $V_\mathrm{L},V_\mathrm{S}=5, 3.5$ (as in Fig.~\ref{fig:CM}a).
In the clean limit, one can clearly see an edge state which connects the lowest energy band with the first excited band.
Therefore the lowest energy gap is topologically non-trivial.
In the strong disorder case there is an intraband state in the lowest energy gap.
This intraband state do not connect the two bands and it is therefore topologically trivial.
Non-trivial edge states are also present in the lowest energy gap in the clean limit $V_\mathrm{D}=0$ for $V_\mathrm{L},V_\mathrm{S}=7.5, 5$, $V_\mathrm{L},V_\mathrm{S}=10, 7$, and $V_\mathrm{L},V_\mathrm{S}=20, 14$.
However, for these choices of the lattice depths, the energy bands dispersion becomes so broad that the lowest energy gaps become very narrow, and the presence and identification of intraband edge states cannot be resolved unambiguously. 

\section{Evaluation of excitation fraction}

In Fig.~3a of the main text, we investigate the gap closing between the first and second bands by measuring the fraction of atoms excited to the second band after three cycles of cRM pumping under quasi-periodic disorder with the lattice parameter $(V_L, V_S)=(20,14)E_R$.
The band structure is defined for the long lattice, spanned by the quasi-momentum $q$,
since we adiabatically turn off the quasi-periodic disorder lattice from the whole lattice setup in 130 ms. We then utilize a band-mapping technique, where an adiabatic turn-off of the remaining optical lattices maps quasi-momentum $q$ of $n$-th band to free-particle momentum $p$ of $n$-th Brillouin zone~\cite{Greiner2001,Kohl2005}.
By taking absorption images after time-of-flight of 10 ms, we evaluate atom fractions in the first and second band from the momentum distribution.
\begin{figure}[t]
\centering
\includegraphics[width=8cm]{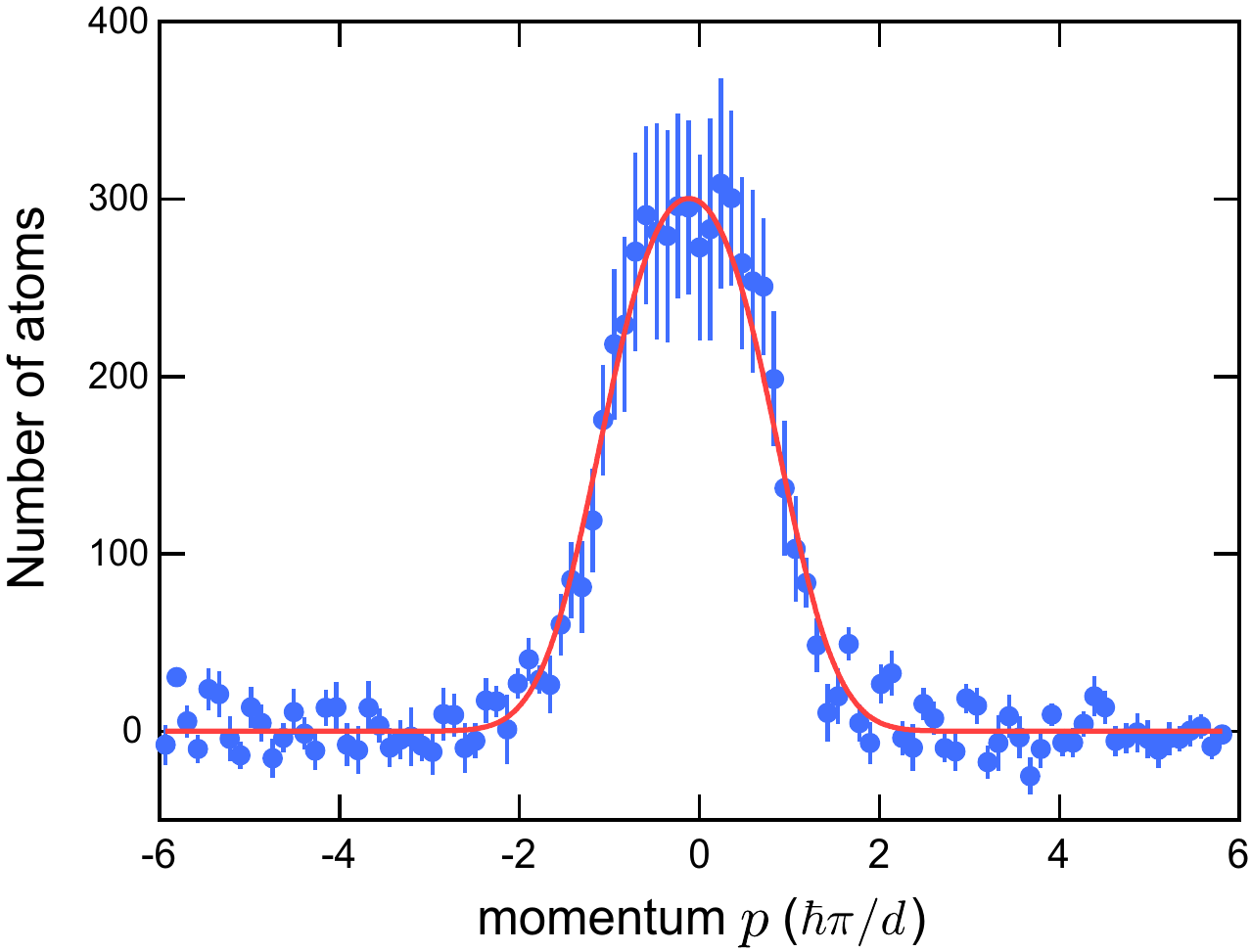}
\caption{
{\bf Measured momentum distribution of atoms in the case of a clean limit with $\phi$ fixed at 0.}
It is fitted with the function  $f_1(p)$ defined in Eq.~(\ref{f1p}).
Error bars denote the standard deviation of ten independent measurements.
}
\label{fig:rb_fit}
\end{figure}

In the case of a clean limit with $\phi$ fixed at 0, it is expected that only the first band is occupied. \figureref{fig:rb_fit} shows the measured momentum distribution corresponding to this case.
An almost homogeneous distribution is obtained within the first Brillouin zone. However, the homogenous distribution was smoothed around $p=\pm \hbar \pi/d$ owing to experimental imperfection such as non-adiabaticity in the band mapping. We approximate this smoothing by the following function $f_1(p)$:
\begin{align}\label{f1p}
f_1(p) = \frac{a}{2}\left[\mathrm{erf}\left(\frac{p+\pi/d}{s}\right)+\mathrm{erf}\left(-\textit{}\frac{p-\pi/d}{s}\right) \right],
\end{align}
here $a$ and $s$ are fitting parameters and the fitted $s$ value $s_\mathrm{fit}$ denotes the smoothness. We assume that this fitted function represents the momentum distribution when the first band is homogeneously occupied.

\begin{figure}[t]
\centering
\includegraphics[width=8cm]{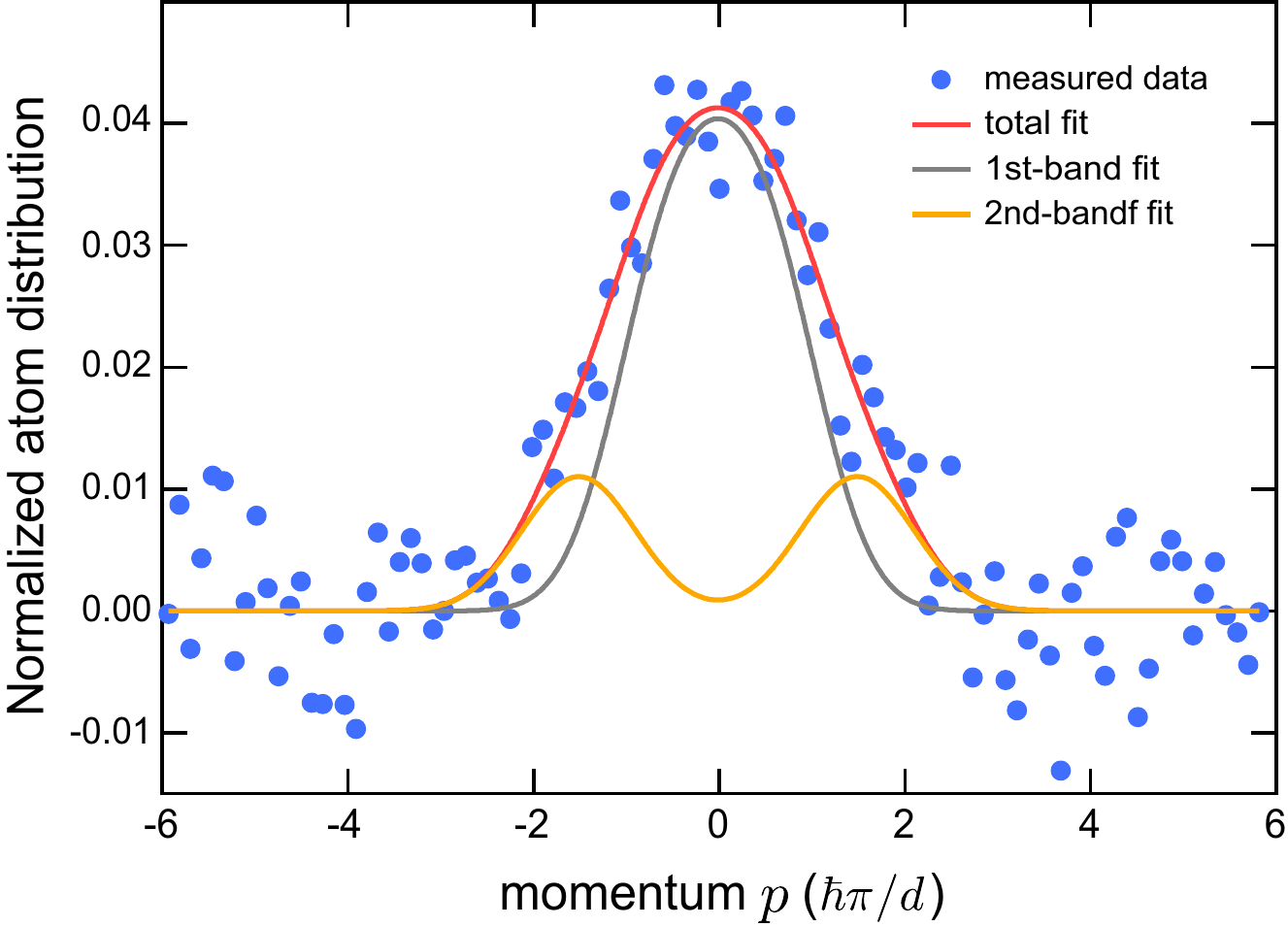}
\caption{
{\bf Measured momentum distribution of atoms after the Thouless pumping under the quasi-periodic disorder.}
This is one typical example of ten independent measurements obtained at the disorder $V_D=20E_R$.
It is fitted with the functions $f_t(p)$, defined in Eq.~(\ref{ftp}), for the whole band (red), $f_1(p)$ for the first band (grey), and $f_2(p)$ for the second band (orange).
The vertical axis is normalized such that the integral of the fitted $f_t(p)$ is unity.
The whole shift in the horizontal axis is corrected such that the functions are symmetric with respect to $p=0$.
}
\label{fig:trb_fit}
\end{figure}

During the pumping under the quasi-periodic disorder, atoms can be excited to the second band. Suppose that the atom distribution in the second band is also homogeneous, and the smoothness of the value $s_\mathrm{fit}$ obtained above is common for both the first and second bands. Then we fit the measured data for the pumping under the quasi-periodic disorder using the following function $f_t(p)$:
\begin{align}\label{ftp}
f_t(p) = f_1(p) + f_2(p),
\end{align}
here
\begin{eqnarray}
f_2(p)=&&\frac{b}{2}\left[\mathrm{erf}\left(\frac{p+2\pi/d}{s_\mathrm{fit}}\right)+\mathrm{erf}\left(-\textit{}\frac{p+\pi/d}{s_\mathrm{fit}}\right) \right] \nonumber \\
&&+\frac{b}{2}\left[\mathrm{erf}\left(\frac{p-\pi/d}{s_\mathrm{fit}}\right)+\mathrm{erf}\left(-\textit{}\frac{p-2\pi/d}{s_\mathrm{fit}}\right) \right],\nonumber \\
\end{eqnarray}
and $b$ is a fitting parameter.
The fitted functions $f_1(p)$ and $f_2(p)$ show the momentum distributions in the first and second bands, respectively.
\figureref{fig:trb_fit} shows one of the fitted results at the disorder $V_D=20E_R$ in Fig.~3a in the main text.
After normalizing the integral of the fitted $f_t(p)$, we evaluated the second-band fraction by calculating $\int dp f_2(p)$.
The mean and error bar representing the standard deviation in Fig.~3a of the main text are obtained for ten independent measurements at each disorder strength.

\section{Intuitive understanding of pump suppression}
We can understand intuitively why the pump is suppressed at $V_D\sim V_L$ in the following way.
As long as the lowest band is occupied, Thouless pumping with a cRM lattice is topologically equivalent to that only with a sliding long lattice~\cite{Nakajima}.
Therefore we can capture the essence of the results of the pump suppression by considering a sliding lattice superimposed with a quasi-periodic disorder lattice, namely the second and third terms of Eq~(1) in the main text. 
\begin{figure*}[t]
\centering
\includegraphics[width=14.5cm]{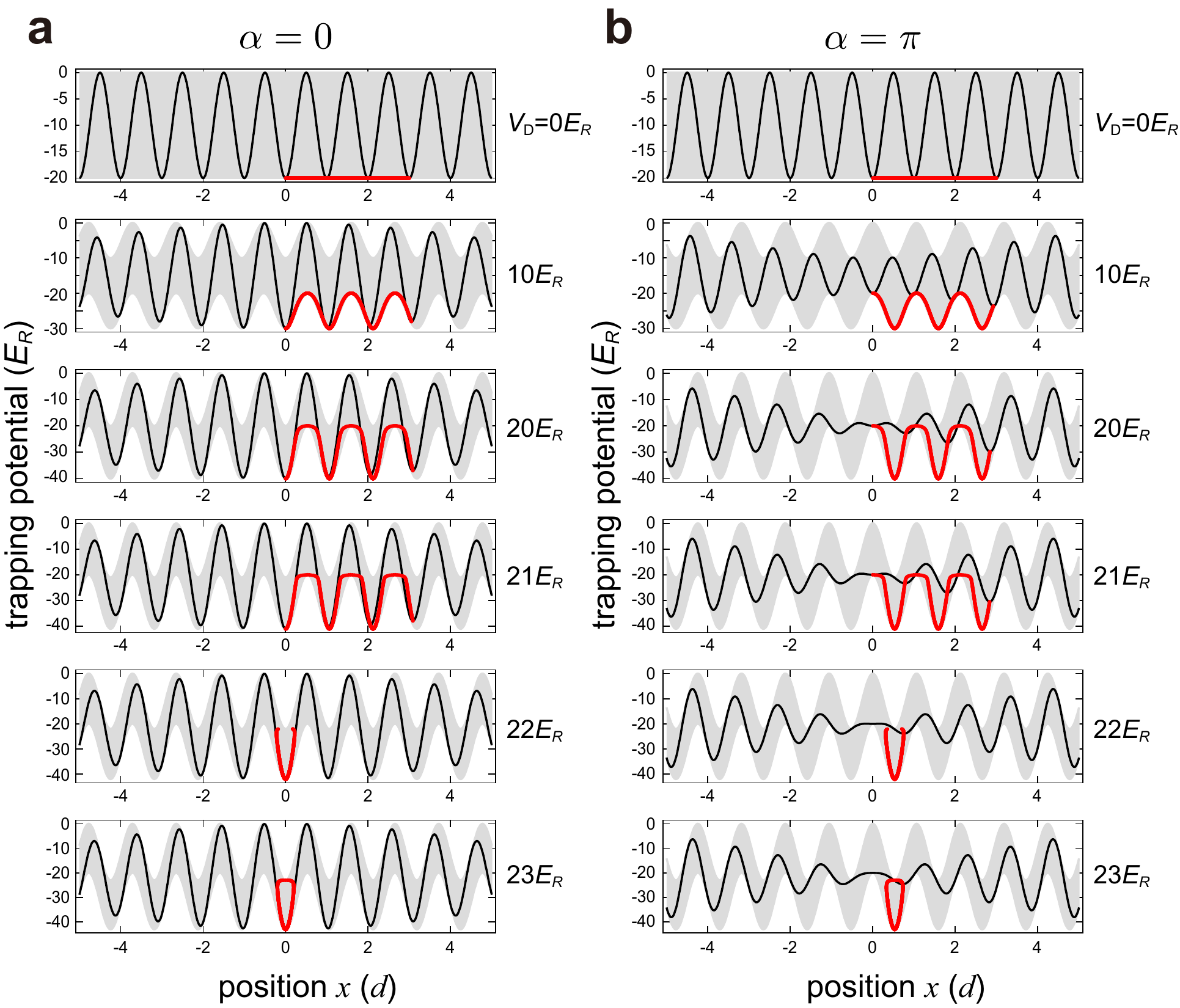}
\caption{
{\bf Position change of the local minimum point in the trapping potential with long and quasi-periodic disorder lattices.}
The long-lattice depth is fixed to $V_L=20E_R$, whereas the disorder strength $V_D$ is changed as described on the right side.
The quasi-periodic disorder phase $\alpha$ is set to 0 ({\bf a}) or $\pi$ ({\bf b}).
The phase $\phi(t)$ is swept from 0 to $3\pi$, corresponding to three pump cycles.
The black line shows the initial potential at $\phi(0)=0$, and the gray shaded area represents the region in which the trapping potential moves during the three cycles.
Position change of the local minimum point of the trap, initially located at $x\approx0$, is depicted as the red line.
}
\label{fig:intuitive}
\end{figure*}

\figureref{fig:intuitive} shows the trapping potential produced by these two lattices as a function of position $x$. In addition, the position change of the local minimum point of the trap during three pump cycles is depicted by the red line.
The trap depth of the long lattice is fixed to $V_L=20E_R$, whereas the disorder strength $V_D$ is varied from $0E_R$ to $23E_R$ (from top to bottom).
When $V_D \leq 21E_R$, the local minimum point moves across several lattice sites during the pump cycle in accordance with the sliding lattice.
This means that the whole lattice is sliding with the phase $\phi$ swept.
For $V_D \geq 22E_R$, however, the minimum point stops moving and wanders around the particular trap minimum.
Accordingly the whole lattice does not slide.
This transition happens around $V_D\sim V_L$.
Therefore it is expected that the pump is suppressed at $V_D\sim V_L$.
Because this discussion does not depend largely on the wavelength of the quasi-periodic disorder lattices, it is reasonable that we observe no clear difference among three measurements in Fig.~4 in the main text.

\begin{figure}[t]
\centering
\includegraphics[width=8cm]{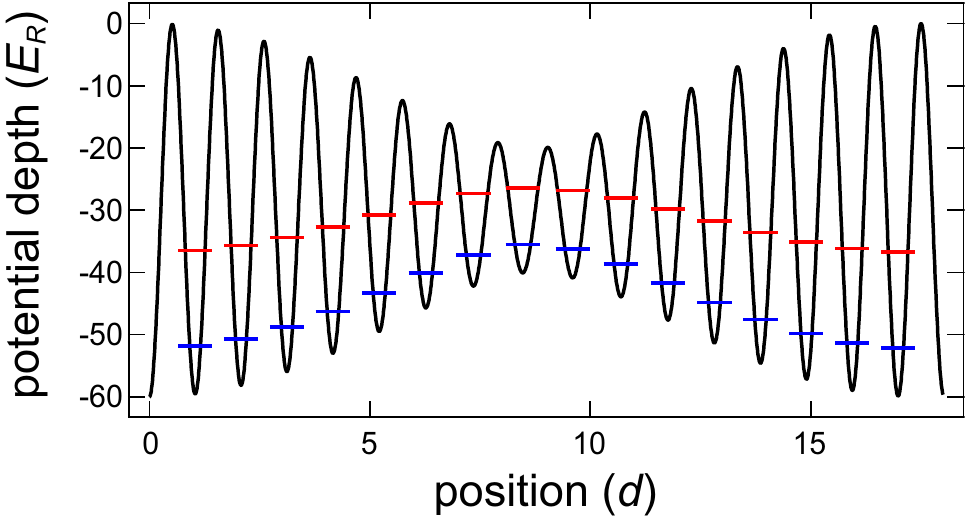}
\caption{
{\bf Trapping potential created by the long and quasi-periodic disorder lattices.} Their trap depths are set to $V_L=20E_R$ and $V_D=40E_R$, respectively, and the relative phase is zero at the origin.
The ground and first-excited vibrational levels are schematically indicated by blue and red bars, respectively.
}
\label{fig:lgap}
\end{figure}

Since the dominant lattice is that of the static quasi-periodic disorder lattice in the case of  $V_D > V_L$, the above-mentioned situation also indicates that the trivial band gap should be open.
Although the re-opening of the gap is not clearly visible in the numerical calculations in Fig.~3 in the main text, this does not necessarily mean that the state is metallic. In fact, the energy difference between the ground and first-excited vibrational levels within individual lattice sites becomes large in the strong disorder as shown in Fig.~\ref{fig:lgap}.
This large local energy gap should suppress the excitations to higher levels as observed in Fig.~3a in the main text.

\begin{thebibliography}{10}
\expandafter\ifx\csname url\endcsname\relax
  \def\url#1{\texttt{#1}}\fi
\expandafter\ifx\csname urlprefix\endcsname\relax\def\urlprefix{URL }\fi
\providecommand{\bibinfo}[2]{#2}
\providecommand{\eprint}[2][]{\url{#2}}

\bibitem{50AL}
\bibinfo{author}{Lagendijk, A.}, \bibinfo{author}{Tiggelen, B.~v.} \&
  \bibinfo{author}{Wiersma, D.~S.}
\newblock \bibinfo{title}{Fifty years of Anderson localization}.
\newblock \emph{\bibinfo{journal}{Physics Today}}
  \textbf{\bibinfo{volume}{62}}, \bibinfo{pages}{24--29} (\bibinfo{year}{2009}).

\bibitem{Niu84}
\bibinfo{author}{Niu, Q.} \& \bibinfo{author}{Thouless, D.~J.}
\newblock \bibinfo{title}{Quantised adiabatic charge transport in the presence
  of substrate disorder and many-body interaction}.
\newblock \emph{\bibinfo{journal}{J. Phys. A: Math. Gen.}}
\textbf{\bibinfo{volume}{17}}, \bibinfo{pages}{2453--2462}
  (\bibinfo{year}{1984}).

\bibitem{Niu85}
\bibinfo{author}{Niu, Q.}, \bibinfo{author}{Thouless, D.~J.} \&
  \bibinfo{author}{Wu, Y.-S.}
\newblock \bibinfo{title}{Quantized Hall conductance as a topological
  invariant}.
\newblock \emph{\bibinfo{journal}{Phys. Rev. B}} \textbf{\bibinfo{volume}{31}},
  \bibinfo{pages}{3372--3377} (\bibinfo{year}{1985}).

\bibitem{Kobayashi}
\bibinfo{author}{Kobayashi, K.}, \bibinfo{author}{Ohtsuki, T.} \&
  \bibinfo{author}{Imura, K.-I.}
\newblock \bibinfo{title}{Disordered weak and strong topological insulators}.
\newblock \emph{\bibinfo{journal}{Phys. Rev. Lett.}}
  \textbf{\bibinfo{volume}{110}}, \bibinfo{pages}{236803}
  (\bibinfo{year}{2013}).

\bibitem{Bryan}
\bibinfo{author}{Leung, B.} \& \bibinfo{author}{Prodan, E.}
\newblock \bibinfo{title}{Effect of strong disorder in a three-dimensional
  topological insulator: Phase diagram and maps of the ${\mathbb{Z}}_{2}$
  invariant}.
\newblock \emph{\bibinfo{journal}{Phys. Rev. B}} \textbf{\bibinfo{volume}{85}},
  \bibinfo{pages}{205136} (\bibinfo{year}{2012}).

\bibitem{Yamakage}
\bibinfo{author}{Yamakage, A.}, \bibinfo{author}{Nomura, K.},
  \bibinfo{author}{Imura, K.-I.} \& \bibinfo{author}{Kuramoto, Y.}
\newblock \bibinfo{title}{Disorder-induced multiple transition involving
  ${\mathbb{Z}}_{2}$ topological insulator}.
\newblock \emph{\bibinfo{journal}{J. Phys. Soc. Jpn.}}
  \textbf{\bibinfo{volume}{80}}, \bibinfo{pages}{053703}
  (\bibinfo{year}{2011}).

\bibitem{Li2009}
\bibinfo{author}{Li, J.}, \bibinfo{author}{Chu, R.-L.}, \bibinfo{author}{Jain,
  J.~K.} \& \bibinfo{author}{Shen, S.-Q.}
\newblock \bibinfo{title}{Topological Anderson insulator}.
\newblock \emph{\bibinfo{journal}{Phys. Rev. Lett.}}
  \textbf{\bibinfo{volume}{102}}, \bibinfo{pages}{136806}
  (\bibinfo{year}{2009}).

\bibitem{McGinley}
\bibinfo{author}{McGinley, M.}, \bibinfo{author}{Knolle, J.} \&
  \bibinfo{author}{Nunnenkamp, A.}
\newblock \bibinfo{title}{Robustness of Majorana edge modes and topological
  order: Exact results for the symmetric interacting Kitaev chain with
  disorder}.
\newblock \emph{\bibinfo{journal}{Phys. Rev. B}} \textbf{\bibinfo{volume}{96}},
  \bibinfo{pages}{241113} (\bibinfo{year}{2017}).

\bibitem{Shen}
\bibinfo{author}{Shen, S.-Q.}
\newblock \emph{\bibinfo{title}{Topological Insulators: Dirac Equation in Condensed Matters}} 
 (\bibinfo{publisher}{Springer}, \bibinfo{year}{2013}).

\bibitem{Meier}
\bibinfo{author}{Meier, E.~J.} \emph{et~al.}
\newblock \bibinfo{title}{Observation of the topological Anderson insulator in
  disordered atomic wires}.
\newblock \emph{\bibinfo{journal}{Science}} \textbf{\bibinfo{volume}{362}},
  \bibinfo{pages}{929--933} (\bibinfo{year}{2018}).

\bibitem{Stutzer2018}
\bibinfo{author}{St{\"{u}}tzer, S.} \emph{et~al.}
\newblock \bibinfo{title}{Photonic topological Anderson insulators}.
\newblock \emph{\bibinfo{journal}{Nature}} \textbf{\bibinfo{volume}{560}},
  \bibinfo{pages}{461--465} (\bibinfo{year}{2018}).

\bibitem{Liu2020}
\bibinfo{author}{Li, G.-G.} \emph{et~al.}
\newblock \bibinfo{title}{Topological Anderson insulator in Disordered Photonic Crystals}.
\newblock \emph{\bibinfo{journal}{Phys. Rev. Lett.}}
  \textbf{\bibinfo{volume}{125}}, \bibinfo{pages}{133603}
  (\bibinfo{year}{2020}).
  
\bibitem{Titum}
\bibinfo{author}{Titum, P.}, \bibinfo{author}{Berg, E.},
  \bibinfo{author}{Rudner, M.~S.}, \bibinfo{author}{Refael, G.} \&
  \bibinfo{author}{Lindner, N.~H.}
\newblock \bibinfo{title}{Anomalous Floquet-Anderson insulator as a
  nonadiabatic quantized charge pump}.
\newblock \emph{\bibinfo{journal}{Phys. Rev. X}} \textbf{\bibinfo{volume}{6}},
  \bibinfo{pages}{021013} (\bibinfo{year}{2016}).

\bibitem{Sriluckshmy}
\bibinfo{author}{Sriluckshmy, P.~V.}, \bibinfo{author}{Saha, K.} \&
  \bibinfo{author}{Moessner, R.}
\newblock \bibinfo{title}{Interplay between topology and disorder in a
  two-dimensional semi-Dirac material}.
\newblock \emph{\bibinfo{journal}{Phys. Rev. B}} \textbf{\bibinfo{volume}{97}},
  \bibinfo{pages}{024204} (\bibinfo{year}{2018}).

\bibitem{Mondragon-Shem}
\bibinfo{author}{Mondragon-Shem, I.} \& \bibinfo{author}{Hughes, T.~L.}
\newblock \bibinfo{title}{Signatures of metal-insulator and topological phase
  transitions in the entanglement of one-dimensional disordered fermions}.
\newblock \emph{\bibinfo{journal}{Phys. Rev. B}} \textbf{\bibinfo{volume}{90}},
  \bibinfo{pages}{104204} (\bibinfo{year}{2014}).

\bibitem{Thouless}
\bibinfo{author}{Thouless, D.~J.}
\newblock \bibinfo{title}{Quantization of particle transport}.
\newblock \emph{\bibinfo{journal}{Phys. Rev. B}}
  \textbf{\bibinfo{volume}{27}}, \bibinfo{pages}{6083--6087}
  (\bibinfo{year}{1983}).

\bibitem{Nakajima}
\bibinfo{author}{Nakajima, S.} \emph{et~al.}
\newblock \bibinfo{title}{Topological Thouless pumping of ultracold fermions}.
\newblock \emph{\bibinfo{journal}{Nat. Phys.}}
  \textbf{\bibinfo{volume}{12}}, \bibinfo{pages}{296--300} (\bibinfo{year}{2016}).

\bibitem{Lohse}
\bibinfo{author}{Lohse, M.}, \bibinfo{author}{Schweizer, C.},
  \bibinfo{author}{Zilberberg, O.}, \bibinfo{author}{Aidelsburger, M.} \&
  \bibinfo{author}{Bloch, I.}
\newblock \bibinfo{title}{A Thouless quantum pump with ultracold bosonic atoms
  in an optical superlattice}.
\newblock \emph{\bibinfo{journal}{Nat. Phys.}}
  \textbf{\bibinfo{volume}{12}}, \bibinfo{pages}{350--354} (\bibinfo{year}{2016}).

\bibitem{Schweizer2016}
\bibinfo{author}{Schweizer, C.}, \bibinfo{author}{Lohse, M.},
  \bibinfo{author}{Citro, R.} \& \bibinfo{author}{Bloch, I.}
\newblock \bibinfo{title}{Spin pumping and measurement of spin currents in
  optical superlattices}.
\newblock \emph{\bibinfo{journal}{Phys. Rev. Lett.}}
  \textbf{\bibinfo{volume}{117}}, \bibinfo{pages}{170405}
  (\bibinfo{year}{2016}).

\bibitem{Lohse2018}
\bibinfo{author}{Lohse, M.}, \bibinfo{author}{Schweizer, C.},
  \bibinfo{author}{Price, H.~M.}, \bibinfo{author}{Zilberberg, O.} \&
  \bibinfo{author}{Bloch, I.}
\newblock \bibinfo{title}{Exploring 4D quantum Hall physics with a 2D
  topological charge pump}.
\newblock \emph{\bibinfo{journal}{Nature}} \textbf{\bibinfo{volume}{553}},
  \bibinfo{pages}{55--58} (\bibinfo{year}{2018}).

\bibitem{Kraus}
\bibinfo{author}{Kraus, Y.~E.}, \bibinfo{author}{Lahini, Y.},
  \bibinfo{author}{Ringel, Z.}, \bibinfo{author}{Verbin, M.} \&
  \bibinfo{author}{Zilberberg, O.}
\newblock \bibinfo{title}{Topological states and adiabatic pumping in
  quasicrystals}.
\newblock \emph{\bibinfo{journal}{Phys. Rev. Lett.}}
  \textbf{\bibinfo{volume}{109}}, \bibinfo{pages}{106402}
  (\bibinfo{year}{2012}).

\bibitem{Zilberberg2018}
\bibinfo{author}{Zilberberg, O.} \emph{et~al.}
\newblock \bibinfo{title}{Photonic topological boundary pumping as a probe of
  4D quantum Hall physics}.
\newblock \emph{\bibinfo{journal}{Nature}} \textbf{\bibinfo{volume}{553}},
  \bibinfo{pages}{59--62} (\bibinfo{year}{2018}).

\bibitem{Qin2016}
\bibinfo{author}{Qin, J.} \& \bibinfo{author}{Guo, H.}
\newblock \bibinfo{title}{Quantum pumping induced by disorder in one
  dimension}.
\newblock \emph{\bibinfo{journal}{Phys. Lett. A}}
  \textbf{\bibinfo{volume}{380}}, \bibinfo{pages}{2317--2321}
  (\bibinfo{year}{2016}).

\bibitem{Wauters}
\bibinfo{author}{Wauters, M.~M.}, \bibinfo{author}{Russomanno, A.},
  \bibinfo{author}{Citro, R.}, \bibinfo{author}{Santoro, G.~E.} \&
  \bibinfo{author}{Privitera, L.}
\newblock \bibinfo{title}{Localization, topology, and quantized transport in
  disordered Floquet systems}.
\newblock \emph{\bibinfo{journal}{Phys. Rev. Lett.}}
  \textbf{\bibinfo{volume}{123}}, \bibinfo{pages}{266601}
  (\bibinfo{year}{2019}).

\bibitem{Imura_2018}
\bibinfo{author}{Imura, K.-I.}, \bibinfo{author}{Yoshimura, Y.},
  \bibinfo{author}{Fukui, T.} \& \bibinfo{author}{Hatsugai, Y.}
\newblock \bibinfo{title}{Bulk-edge correspondence in topological transport and
  pumping}.
\newblock \emph{\bibinfo{journal}{J. Phys. Conf. Ser.}}
  \textbf{\bibinfo{volume}{969}}, \bibinfo{pages}{012133}
  (\bibinfo{year}{2018}).

\bibitem{Kuno}
\bibinfo{author}{Kuno, Y.}
\newblock \bibinfo{title}{Disorder-induced Chern insulator in the
  Harper-Hofstadter-Hatsugai model}.
\newblock \emph{\bibinfo{journal}{Phys. Rev. B}}
  \textbf{\bibinfo{volume}{100}}, \bibinfo{pages}{054108}
  (\bibinfo{year}{2019}).

\bibitem{Nakagawa}
\bibinfo{author}{Nakagawa, M.}, \bibinfo{author}{Yoshida, T.},
  \bibinfo{author}{Peters, R.} \& \bibinfo{author}{Kawakami, N.}
\newblock \bibinfo{title}{Breakdown of topological Thouless pumping in the
  strongly interacting regime}.
\newblock \emph{\bibinfo{journal}{Phys. Rev. B}} \textbf{\bibinfo{volume}{98}},
  \bibinfo{pages}{115147} (\bibinfo{year}{2018}).

\bibitem{Hayward}
\bibinfo{author}{Hayward, A.}, \bibinfo{author}{Schweizer, C.},
  \bibinfo{author}{Lohse, M.}, \bibinfo{author}{Aidelsburger, M.} \&
  \bibinfo{author}{Heidrich-Meisner, F.}
\newblock \bibinfo{title}{Topological charge pumping in the interacting bosonic
  Rice-Mele model}.
\newblock \emph{\bibinfo{journal}{Phys. Rev. B}} \textbf{\bibinfo{volume}{98}},
  \bibinfo{pages}{245148} (\bibinfo{year}{2018}).

\bibitem{Stenzel}
\bibinfo{author}{Stenzel, L.}, \bibinfo{author}{Hayward, A. L.~C.},
  \bibinfo{author}{Hubig, C.}, \bibinfo{author}{Schollw\"ock, U.} \&
  \bibinfo{author}{Heidrich-Meisner, F.}
\newblock \bibinfo{title}{Quantum phases and topological properties of
  interacting fermions in one-dimensional superlattices}.
\newblock \emph{\bibinfo{journal}{Phys. Rev. A}} \textbf{\bibinfo{volume}{99}},
  \bibinfo{pages}{053614} (\bibinfo{year}{2019}).

\bibitem{Mei_2019}
\bibinfo{author}{Mei, F.}, \bibinfo{author}{Chen, G.},
  \bibinfo{author}{Goldman, N.}, \bibinfo{author}{Xiao, L.} \&
  \bibinfo{author}{Jia, S.}
\newblock \bibinfo{title}{Topological magnon insulator and quantized pumps from
  strongly-interacting bosons in optical superlattices}.
\newblock \emph{\bibinfo{journal}{New J. Phys.}}
  \textbf{\bibinfo{volume}{21}}, \bibinfo{pages}{095002}
  (\bibinfo{year}{2019}).

\bibitem{RM}
\bibinfo{author}{Rice, M.~J.} \& \bibinfo{author}{Mele, E.~J.}
\newblock \bibinfo{title}{Elementary excitations of a linearly conjugated
  diatomic polymer}.
\newblock \emph{\bibinfo{journal}{Phys. Rev. Lett.}}
  \textbf{\bibinfo{volume}{49}}, \bibinfo{pages}{1455--1459}
  (\bibinfo{year}{1982}).

\bibitem{Atala}
\bibinfo{author}{Atala, M.} \emph{et~al.}
\newblock \bibinfo{title}{Direct measurement of the Zak phase in topological
  Bloch bands}.
\newblock \emph{\bibinfo{journal}{Nat. Phys.}} \textbf{\bibinfo{volume}{9}},
  \bibinfo{pages}{795--800} (\bibinfo{year}{2013}).

\bibitem{Hayward2020}
\bibinfo{author}{Hayward, A. L. C.}, \bibinfo{author}{Bertok, E.},
  \bibinfo{author}{Schneider, U.} \& \bibinfo{author}{Heidrich-Meisner, F.}
\newblock \bibinfo{title}{Effect of disorder on topological charge pumping in the Rice-Mele model}.
\newblock \emph{\bibinfo{journal}{Phys. Rev. A}} \textbf{\bibinfo{volume}{103}},
  \bibinfo{pages}{043310} (\bibinfo{year}{2021}).

\bibitem{Asboth}
\bibinfo{author}{Asb{\'o}th, J.~K.}, \bibinfo{author}{Oroszl{\'a}ny, L.} \&
  \bibinfo{author}{P{\'a}lyi, A.}
\newblock \bibinfo{title}{A short course on topological insulators}.
\newblock \emph{\bibinfo{journal}{Lecture notes in physics}}
  \textbf{\bibinfo{volume}{919}} (\bibinfo{year}{2016}).

\bibitem{YFZhang}
\bibinfo{author}{Zhang, Y.-F.} \emph{et~al.}
\newblock \bibinfo{title}{Coupling-matrix approach to the Chern number
  calculation in disordered systems}.
\newblock \emph{\bibinfo{journal}{Chin. Phys. B}}
  \textbf{\bibinfo{volume}{22}}, \bibinfo{pages}{117312}
  (\bibinfo{year}{2013}).

\bibitem{Jotzu14}
\bibinfo{author}{Jotzu, G.} \emph{et~al.}
\newblock \bibinfo{title}{Experimental realization of the topological Haldane
  model with ultracold fermions}.
\newblock \emph{\bibinfo{journal}{Nature}} \textbf{\bibinfo{volume}{515}},
  \bibinfo{pages}{237--240} (\bibinfo{year}{2014}).

\bibitem{Das2019}
\bibinfo{author}{Das, K.~K.} \& \bibinfo{author}{Christ, J.}
\newblock \bibinfo{title}{Realizing the Harper model with ultracold atoms in a
  ring lattice}.
\newblock \emph{\bibinfo{journal}{Phys. Rev. A}} \textbf{\bibinfo{volume}{99}},
  \bibinfo{pages}{013604} (\bibinfo{year}{2019}).

\bibitem{Marra2020}
\bibinfo{author}{{Marra}, P.} \& \bibinfo{author}{{Nitta}, M.}
\newblock \bibinfo{title}{{Topologically quantized current in quasiperiodic
  Thouless pumps}}.
\newblock \emph{\bibinfo{journal}{Phys. Rev. Research}} \textbf{\bibinfo{volume}{2}},
  \bibinfo{pages}{042035} (\bibinfo{year}{2020}).

\bibitem{AAmodel}
\bibinfo{author}{{Aubry}, S.} \& \bibinfo{author}{{Andr{\'e}}, G.}
\newblock \bibinfo{title}{Analyticity breaking and Anderson localization in incommensurate lattices}.
\newblock \emph{\bibinfo{journal}{Ann. Israel Phys. Soc.}} \textbf{\bibinfo{volume}{3}},
  \bibinfo{pages}{133--164} (\bibinfo{year}{1980}).

\bibitem{Huse}
\bibinfo{author}{Huse, D.~A.}, \bibinfo{author}{Nandkishore, R.},
  \bibinfo{author}{Oganesyan, V.}, \bibinfo{author}{Pal, A.} \&
  \bibinfo{author}{Sondhi, S.~L.}
\newblock \bibinfo{title}{Localization-protected quantum order}.
\newblock \emph{\bibinfo{journal}{Phys. Rev. B}} \textbf{\bibinfo{volume}{88}},
  \bibinfo{pages}{014206} (\bibinfo{year}{2013}).

\bibitem{Nandkishore}
\bibinfo{author}{Nandkishore, R.} \& \bibinfo{author}{Huse, D.~A.}
\newblock \bibinfo{title}{Many-body localization and thermalization in quantum
  statistical mechanics}.
\newblock \emph{\bibinfo{journal}{Annu. Rev. Condens. Matter Phys.}}
  \textbf{\bibinfo{volume}{6}}, \bibinfo{pages}{15--38} (\bibinfo{year}{2015}).

\bibitem{Price2015}
\bibinfo{author}{Price, H.~M.}, \bibinfo{author}{Zilberberg, O.},
  \bibinfo{author}{Ozawa, T.}, \bibinfo{author}{Carusotto, I.} \&
  \bibinfo{author}{Goldman, N.}
\newblock \bibinfo{title}{Four-dimensional quantum Hall effect with ultracold
  atoms}.
\newblock \emph{\bibinfo{journal}{Phys. Rev. Lett.}}
  \textbf{\bibinfo{volume}{115}}, \bibinfo{pages}{195303}
  (\bibinfo{year}{2015}).

\bibitem{Ippoliti2020}
\bibinfo{author}{Ippoliti, M.} \& \bibinfo{author}{Bhatt, R.~N.}
\newblock \bibinfo{title}{Dimensional crossover of the integer quantum Hall
  plateau transition and disordered topological pumping}.
\newblock \emph{\bibinfo{journal}{Phys. Rev. Lett.}}
  \textbf{\bibinfo{volume}{124}}, \bibinfo{pages}{086602}
  (\bibinfo{year}{2020}).

\bibitem{Citro2018}
\bibinfo{author}{Privitera, L.}, \bibinfo{author}{Russomanno, A.}, \bibinfo{author}{Citro, R.} \& 
\bibinfo{author}{Santoro, G.~E.}
\newblock \bibinfo{title}{Nonadiabatic Breaking of Topological Pumping}.
\newblock \emph{\bibinfo{journal}{Phys. Rev. Lett.}}
  \textbf{\bibinfo{volume}{120}}, \bibinfo{pages}{106601}
  (\bibinfo{year}{2018}).

\bibitem{Benalcazar61}
\bibinfo{author}{Benalcazar, W.~A.}, \bibinfo{author}{Bernevig, B.~A.} \&
  \bibinfo{author}{Hughes, T.~L.}
\newblock \bibinfo{title}{Quantized electric multipole insulators}.
\newblock \emph{\bibinfo{journal}{Science}} \textbf{\bibinfo{volume}{357}},
  \bibinfo{pages}{61--66} (\bibinfo{year}{2017}).

\bibitem{Xie}
\bibinfo{author}{Xie, B.-Y.} \emph{et~al.}
\newblock \bibinfo{title}{Second-order photonic topological insulator with
  corner states}.
\newblock \emph{\bibinfo{journal}{Phys. Rev. B}} \textbf{\bibinfo{volume}{98}},
  \bibinfo{pages}{205147} (\bibinfo{year}{2018}).

\bibitem{SQShen}
\bibinfo{author}{Li, C.-A.}, {Fu, B.}, {Hu, Z.-A.}, {Li, J.} \& {Shen, S.-Q.}
\newblock \bibinfo{title}{Topological Phase Transitions in Disordered Electric Quadrupole Insulators}.
\newblock \emph{\bibinfo{journal}{Phys. Rev. Lett.}}
  \textbf{\bibinfo{volume}{125}}, \bibinfo{pages}{166801}
  (\bibinfo{year}{2020}).

\bibitem{Araki}
\bibinfo{author}{Araki, H.}, \bibinfo{author}{Mizoguchi, T.} \&
  \bibinfo{author}{Hatsugai, Y.}
\newblock \bibinfo{title}{Phase diagram of a disordered higher-order
  topological insulator: A machine learning study}.
\newblock \emph{\bibinfo{journal}{Phys. Rev. B}} \textbf{\bibinfo{volume}{99}},
  \bibinfo{pages}{085406} (\bibinfo{year}{2019}).

\bibitem{Kitagawa08}
\bibinfo{author}{Kitagawa, M.} \emph{et~al.}
\newblock \bibinfo{title}{Two-color photoassociation spectroscopy of ytterbium
  atoms and the precise determinations of s-wave scattering lengths}.
\newblock \emph{\bibinfo{journal}{Phys. Rev. A}}
  \textbf{\bibinfo{volume}{77}}, \bibinfo{pages}{012719}
  (\bibinfo{year}{2008}).

\bibitem{Taie10}
\bibinfo{author}{Taie, S.} \emph{et~al.}
\newblock \bibinfo{title}{Realization of a SU(2) $\times$ SU(6) system of fermions in
  a cold atomic gas}.
\newblock \emph{\bibinfo{journal}{Phys. Rev. Lett.}}
  \textbf{\bibinfo{volume}{105}}, \bibinfo{pages}{190401}
  (\bibinfo{year}{2010}).

\end{thebibliography}

\begin{thebibliography}{10}
\expandafter\ifx\csname url\endcsname\relax
  \def\url#1{\texttt{#1}}\fi
\expandafter\ifx\csname urlprefix\endcsname\relax\def\urlprefix{URL }\fi
\providecommand{\bibinfo}[2]{#2}
\providecommand{\eprint}[2][]{\url{#2}}

\bibitem{Nakajima}
\bibinfo{author}{Nakajima, S.} \emph{et~al.}
\newblock \bibinfo{title}{Topological thouless pumping of ultracold fermions}.
\newblock \emph{\bibinfo{journal}{Nature Physics}}
  \textbf{\bibinfo{volume}{12}}, \bibinfo{pages}{296} (\bibinfo{year}{2016}).

\bibitem{Schnyder}
\bibinfo{author}{Schnyder, A.~P.}, \bibinfo{author}{Ryu, S.},
  \bibinfo{author}{Furusaki, A.} \& \bibinfo{author}{Ludwig, A. W.~W.}
\newblock \bibinfo{title}{Classification of topological insulators and
  superconductors in three spatial dimensions}.
\newblock \emph{\bibinfo{journal}{Phys. Rev. B}} \textbf{\bibinfo{volume}{78}},
  \bibinfo{pages}{195125} (\bibinfo{year}{2008}).

\bibitem{Kitaev}
\bibinfo{author}{Kitaev, A.}
\newblock \bibinfo{title}{Periodic table for topological insulators and
  superconductors}.
\newblock \emph{\bibinfo{journal}{AIP Conference Proceedings}}
  \textbf{\bibinfo{volume}{1134}}, \bibinfo{pages}{22--30}
  (\bibinfo{year}{2009}).

\bibitem{Asboth}
\bibinfo{author}{Asb{\'o}th, J.~K.}, \bibinfo{author}{Oroszl{\'a}ny, L.} \&
  \bibinfo{author}{P{\'a}lyi, A.}
\newblock \bibinfo{title}{A short course on topological insulators}.
\newblock \emph{\bibinfo{journal}{Lecture notes in physics}}
  \textbf{\bibinfo{volume}{919}} (\bibinfo{year}{2016}).

\bibitem{Meier}
\bibinfo{author}{Meier, E.~J.} \emph{et~al.}
\newblock \bibinfo{title}{Observation of the topological anderson insulator in
  disordered atomic wires}.
\newblock \emph{\bibinfo{journal}{Science}} \textbf{\bibinfo{volume}{362}},
  \bibinfo{pages}{929--933} (\bibinfo{year}{2018}).

\bibitem{Wauters}
\bibinfo{author}{Wauters, M.~M.}, \bibinfo{author}{Russomanno, A.},
  \bibinfo{author}{Citro, R.}, \bibinfo{author}{Santoro, G.~E.} \&
  \bibinfo{author}{Privitera, L.}
\newblock \bibinfo{title}{Localization, topology, and quantized transport in
  disordered Floquet systems}.
\newblock \emph{\bibinfo{journal}{Phys. Rev. Lett.}}
  \textbf{\bibinfo{volume}{123}}, \bibinfo{pages}{266601}
  (\bibinfo{year}{2019}).

\bibitem{Hatsugai}
\bibinfo{author}{Hatsugai, Y.} \& \bibinfo{author}{Kohmoto, M.}
\newblock \bibinfo{title}{Energy spectrum and the quantum hall effect on the
  square lattice with next-nearest-neighbor hopping}.
\newblock \emph{\bibinfo{journal}{Phys. Rev. B}} \textbf{\bibinfo{volume}{42}},
  \bibinfo{pages}{8282--8294} (\bibinfo{year}{1990}).

\bibitem{Kraus}
\bibinfo{author}{Kraus, Y.~E.}, \bibinfo{author}{Lahini, Y.},
  \bibinfo{author}{Ringel, Z.}, \bibinfo{author}{Verbin, M.} \&
  \bibinfo{author}{Zilberberg, O.}
\newblock \bibinfo{title}{Topological states and adiabatic pumping in
  quasicrystals}.
\newblock \emph{\bibinfo{journal}{Phys. Rev. Lett.}}
  \textbf{\bibinfo{volume}{109}}, \bibinfo{pages}{106402}
  (\bibinfo{year}{2012}).

\bibitem{Lohse}
\bibinfo{author}{Lohse, M.}, \bibinfo{author}{Schweizer, C.},
  \bibinfo{author}{Zilberberg, O.}, \bibinfo{author}{Aidelsburger, M.} \&
  \bibinfo{author}{Bloch, I.}
\newblock \bibinfo{title}{A thouless quantum pump with ultracold bosonic atoms
  in an optical superlattice}.
\newblock \emph{\bibinfo{journal}{Nature Physics}}
  \textbf{\bibinfo{volume}{12}}, \bibinfo{pages}{350} (\bibinfo{year}{2016}).

\bibitem{Hofstadter}
\bibinfo{author}{Hofstadter, D.~R.}
\newblock \bibinfo{title}{Energy levels and wave functions of bloch electrons
  in rational and irrational magnetic fields}.
\newblock \emph{\bibinfo{journal}{Phys. Rev. B}} \textbf{\bibinfo{volume}{14}},
  \bibinfo{pages}{2239--2249} (\bibinfo{year}{1976}).

\bibitem{YFZhang}
\bibinfo{author}{Zhang, Y.-F.} \emph{et~al.}
\newblock \bibinfo{title}{Coupling-matrix approach to the chern number
  calculation in disordered systems}.
\newblock \emph{\bibinfo{journal}{Chinese Physics B}}
  \textbf{\bibinfo{volume}{22}}, \bibinfo{pages}{117312}
  (\bibinfo{year}{2013}).

\bibitem{Castro}
\bibinfo{author}{Castro, E.~V.}, \bibinfo{author}{L\'opez-Sancho, M.~P.} \&
  \bibinfo{author}{Vozmediano, M. A.~H.}
\newblock \bibinfo{title}{Anderson localization and topological transition in
  chern insulators}.
\newblock \emph{\bibinfo{journal}{Phys. Rev. B}} \textbf{\bibinfo{volume}{92}},
  \bibinfo{pages}{085410} (\bibinfo{year}{2015}).

\bibitem{Sriluckshmy}
\bibinfo{author}{Sriluckshmy, P.~V.}, \bibinfo{author}{Saha, K.} \&
  \bibinfo{author}{Moessner, R.}
\newblock \bibinfo{title}{Interplay between topology and disorder in a
  two-dimensional semi-dirac material}.
\newblock \emph{\bibinfo{journal}{Phys. Rev. B}} \textbf{\bibinfo{volume}{97}},
  \bibinfo{pages}{024204} (\bibinfo{year}{2018}).

\bibitem{Kuno}
\bibinfo{author}{Kuno, Y.}
\newblock \bibinfo{title}{Disorder-induced chern insulator in the
  harper-hofstadter-hatsugai model}.
\newblock \emph{\bibinfo{journal}{Phys. Rev. B}}
  \textbf{\bibinfo{volume}{100}}, \bibinfo{pages}{054108}
  (\bibinfo{year}{2019}).

\bibitem{Niu85}
\bibinfo{author}{Niu, Q.}, \bibinfo{author}{Thouless, D.~J.} \&
  \bibinfo{author}{Wu, Y.-S.}
\newblock \bibinfo{title}{Quantized hall conductance as a topological
  invariant}.
\newblock \emph{\bibinfo{journal}{Phys. Rev. B}} \textbf{\bibinfo{volume}{31}},
  \bibinfo{pages}{3372--3377} (\bibinfo{year}{1985}).

\bibitem{continued_fractions_hardy}
\bibinfo{author}{Hardy, G.~H.} \& \bibinfo{author}{Wright, E.~M.}
\newblock \emph{\bibinfo{title}{An Introduction to the Theory of Numbers}}
  (\bibinfo{publisher}{Oxford University Press}, \bibinfo{address}{Oxford},
  \bibinfo{year}{2008}), \bibinfo{edition}{6} edn.
\newblock \bibinfo{note}{See chapters 10 and 11}.

\bibitem{Greiner2001}
\bibinfo{author}{Greiner, M.}, \bibinfo{author}{Bloch, I.},
  \bibinfo{author}{Mandel, O.}, \bibinfo{author}{H\"ansch, T.~W.} \&
  \bibinfo{author}{Esslinger, T.}
\newblock \bibinfo{title}{Exploring phase coherence in a 2d lattice of
  bose-einstein condensates}.
\newblock \emph{\bibinfo{journal}{Phys. Rev. Lett.}}
  \textbf{\bibinfo{volume}{87}}, \bibinfo{pages}{160405}
  (\bibinfo{year}{2001}).

\bibitem{Kohl2005}
\bibinfo{author}{K\"ohl, M.}, \bibinfo{author}{Moritz, H.},
  \bibinfo{author}{St\"oferle, T.}, \bibinfo{author}{G\"unter, K.} \&
  \bibinfo{author}{Esslinger, T.}
\newblock \bibinfo{title}{Fermionic atoms in a three dimensional optical
  lattice: Observing fermi surfaces, dynamics, and interactions}.
\newblock \emph{\bibinfo{journal}{Phys. Rev. Lett.}}
  \textbf{\bibinfo{volume}{94}}, \bibinfo{pages}{080403}
  (\bibinfo{year}{2005}).

\end{thebibliography}

\providecommand{\noopsort}[1]{}\providecommand{\singleletter}[1]{#1}%

\clearpage

\end{document}